\documentclass[prd,twocolumn,groupedaddress,english,floatfix,superscriptaddress,preprintnumbers,nofootinbib]{revtex4-1}
\usepackage{graphics}
\usepackage{cancel}
\usepackage{amssymb}
\usepackage{textcomp}
\usepackage{amsmath}
\usepackage{mathtools}
\usepackage{bm}
\usepackage{times}
\usepackage{epsfig}
\usepackage[dvipsnames]{xcolor}
\usepackage{hyperref}
\usepackage{setspace}
\usepackage{comment}
\usepackage{subcaption}
\usepackage[utf8]{inputenc}
\usepackage[english]{babel}
\usepackage{textcomp}
\usepackage{latexsym}
\usepackage{mathrsfs}
\usepackage[section]{placeins}
\usepackage{cleveref}

\usepackage[normalem]{ulem}

\usepackage{tikz}
\usetikzlibrary{patterns}
\usetikzlibrary{shapes.geometric}
\usetikzlibrary{decorations.pathmorphing}
\usetikzlibrary{arrows.meta}
\tikzset{snake it/.style={decorate, decoration=snake}}
\usetikzlibrary{positioning}
\usetikzlibrary{arrows,shapes,positioning}
\usetikzlibrary{decorations.markings}
\tikzstyle arrowstyle=[scale=1]
\tikzstyle directed=[postaction={decorate,decoration={markings,mark=at position .65 with {\arrow[arrowstyle]{stealth}}}}]
\tikzstyle reverse directed=[postaction={decorate,decoration={markings,mark=at position .65 with {\arrowreversed[arrowstyle]{stealth};}}}]

\tikzset{->-/.style={decoration={
  markings,
  mark=at position #1 with {\arrow{>}}},postaction={decorate}}}

\tikzset{-<-/.style={decoration={
  markings,
  mark=at position #1 with {\arrow{<}}},postaction={decorate}}}

\newcommand{\bR}{\ensuremath{\mathbb{R}}}
\newcommand{\bS}{\ensuremath{\mathbb{S}}}

\newcommand{\rC}{{\mathrm{C}}}
\newcommand{\rU}{{\mathrm{U}}}
\newcommand{\rI}{{\mathrm{I}}}
\newcommand{\rII}{{\mathrm{II}}}
\newcommand{\rIII}{{\mathrm{III}}}

\newcommand{\shorter}[1]{\textcolor{green}{}}

\newcommand{\marcc}[1]{}

\newcommand{\mH}{\ensuremath{\mathcal{H}}}

\newcommand{\ri}{\ensuremath{\mathrm{in}}}
\newcommand{\ru}{\ensuremath{\mathrm{up}}}

\newcommand{\VEV}[1]{{\left\langle #1 \right\rangle}}
\newcommand{\abs}[1]{{\left\vert #1 \right\vert}}
\newcommand{\del}{{\partial}}
\newcommand{\td}{\ensuremath{\,\text{d}}}
\newcommand{\vp}{\ensuremath{\varphi}}

\renewcommand{\Re}{{\rm Re}}

\DeclareMathOperator{\csch}{csch}

\begin{document}
\title{Infinite quantum twisting at the Cauchy horizon of rotating black holes}
\author{Christiane Klein}
\email{christiane.klein@univ-grenoble-alpes.fr}
\affiliation{Institut f\"ur Theoretische Physik, Universit\"at Leipzig,\\ Br\"uderstra{\ss}e 16, 04103 Leipzig, Germany}
\affiliation{Univ. Grenoble Alpes, CNRS, IF, 38000 Grenoble, France}
\affiliation{AGM, CY Cergy Paris Université, 2 av. Adolphe Chauvin 95302 Cergy-Pontoise, France}
\author{Mojgan Soltani}
\email{mojgan.soltani@cfel.de}
\affiliation{Institut f\"ur Theoretische Physik, Universit\"at Leipzig,\\ Br\"uderstra{\ss}e 16, 04103 Leipzig, Germany}
\affiliation{DESY, Notkestraße 85, 22607 Hamburg, Germany}
\author{Marc Casals}
\email{marc.casals@uni-leipzig.de}
\affiliation{Institut f\"ur Theoretische Physik, Universit\"at Leipzig,\\ Br\"uderstra{\ss}e 16, 04103 Leipzig, Germany}
\affiliation{School of Mathematics and Statistics, University College Dublin, Belfield, Dublin 4, Ireland.}
\affiliation{Centro Brasileiro de Pesquisas F\'isicas (CBPF), Rio de Janeiro, 
CEP 22290-180, 
Brazil.}
\author{Stefan Hollands}
\email{stefan.hollands@uni-leipzig.de}
\affiliation{Institut f\"ur Theoretische Physik, Universit\"at Leipzig,\\ Br\"uderstra{\ss}e 16, 04103 Leipzig, Germany}
\affiliation{MPI-MiS, Inselstrasse 22, 04103 Leipzig, Germany}

\begin{abstract}
We present a numerical calculation of the expectation value of the quantum angular-momentum current flux density for a scalar field in the Unruh state near the inner horizon of a Kerr-de Sitter black hole. Our results indicate that this flux diverges as $V_-^{-1}$ in a suitable 
    Kruskal coordinate such that $V_-=0$ at the inner horizon. Depending on the parameter values of the scalar field and black hole that we consider, and depending on the polar angle (latitude), this flux can have different signs. In the near extremal cases considered, the angle average of the expectation value of the quantum angular momentum current flux is of the opposite sign as the angular momentum of the background itself, suggesting that, in the cases considered, quantum effects 
     tend to decrease the total angular
momentum of the spheres away from the extremal value. 
     We also  numerically calculate the energy flux component, which  provides the leading order divergence of the quantum stress energy tensor, dominant over the classical stress energy tensor, at the inner horizon. Taking our expectation value of the quantum stress tensor as the source in the semiclassical Einstein equation, our analysis suggests that the spheres approaching the inner horizon can undergo an infinite twisting due to quantum effects   along latitudes separating regions of infinite expansion and contraction.
\end{abstract}

\maketitle


\section{Introduction}
\label{sec:intro}

Black hole (BH) spacetimes with non-vanishing charge and/or angular momentum possess inner horizons (IHs). These are Cauchy horizons  because they delineate the domain of dependence of the initial value problem for classical or quantum wave-type equations posed on an initial Cauchy surface. 

Penrose \cite{Penrose:1974} suggested that perturbations of the BH metric, as well as matter fields on these spacetimes, might diverge at the IH. If his 
``strong cosmic censorship hypothesis'' (sCC)  were true, then the regular IHs present in these exact solutions of Einstein equations would be mere mathematical illusions and should get converted to  a non-timelike singularity in a full (classical) treatment of the coupled Einstein-matter system. Such a scenario would be an elegant resolution of the issues of determinism raised by IHs, relegating them  to the realm of quantum gravity taking over sufficiently near these singularities. 

The validity of sCC has been investigated in many works, starting 
with \cite{Poisson:1989,Poisson:1990, Ori:1991} who  accumulated suggestive evidence in favor of the hypothesis. One question concerns the exact nature of the singularity. Recent mathematical works on the full Einstein equations \cite{Dafermos:2017} show that tidal distortions at the singularity remain finite for sufficiently small initial perturbations \footnote{For essentially  nonlinear effects in spherical symmetry, see \cite{li2023kasner}.}, but one expects that tidal forces suitably diverge in such a way as to make the metric inextensible as a {\it weak} solution to the Einstein equations. According to a formulation by \cite{Christodoulou:2008}, this means that first derivatives of the metric ought to fail to be locally square integrable at the IH. In terms of scalar test-fields, 
it means that some component $T_{\mu\nu}$ of the matter stress-energy should fail to be locally integrable there.
Although a proof of this strong form of sCC is still missing in the context of the full Einstein-matter equations, there is a large body of evidence in the context of the test field approximation of the Einstein equations, and via numerical approaches \cite{Mellor:1990, Mellor:1992, Brady:1998, Dafermos:2015, Luk:2015, Cardoso:2018, Dias:2018, Dias:2018a, Costa:2018, Luna:2019}. 

The charge and angular momentum of BHs have upper (extremal) bounds.
 Interestingly, it has recently been observed~\cite{Cardoso:2017, Cardoso:2018, Dias:2018} that sCC is classically violated near the extremal limit of
charged, static BHs (Reissner-Nordström, RN) in a {\it de Sitter} Universe (i.e., with a positive cosmological constant $\Lambda$), RNdS. Thus, one naturally wonders whether quantum effects could become relevant in such a setting. It was indeed shown by \cite{Hollands:2019,Hollands:2020} (building partly on earlier pioneering work by \cite{Lanir:2017,Zilberman:2019}) that the component of the renormalized expectation value $\langle \hat T_{\mu\nu} \rangle$  of the quantum stress-energy tensor (RSET) which is relevant for the shear and expansion of a congruence of observers crossing the IH of RNdS has a quadratic divergence in a Kruskal (regular)
coordinate. The leading order divergence is independent of the chosen initial Hadamard, i.e. `regular', state and entirely of quantum nature, in the sense that the difference between the RSETs in two states diverges at a (weaker) rate set by the classical theory. 
 
 The nature of quantum effects 
at the IH cannot be naively explained by  spontaneous pair production from the vacuum: 
a charged scalar quantum field would always discharge the IH~\cite{Herman:1994, Sorkin:2000} according to Schwinger pair creation~\cite{Schwinger:1951}, whereas
a full quantum field theory computation \cite{Klein:2021, Klein:2021a} of the expectation value of the charge current revealed that the current may increase the charge of the IH of RNdS in a certain parameter region (although near extremality, the charge is decreased).

Charged BHs can be considered as toy models for the more complicated, and astrophysically-relevant, rotating ones: Kerr when $\Lambda=0$ and Kerr-de Sitter (KdS) when $\Lambda>0$. 
Even though in Kerr and KdS, sCC is expected to hold already at the classical level~\cite{Ori:1991,Dias:2018a} \footnote{sCC also holds classically in RN but not in rotating and charged BHs in de Sitter~\cite{2022PhRvD.106d4060C}.}, the calculation of quantum fluxes is nevertheless interesting because one would like to see whether they can dominate and/or be qualitatively different at the IH even in such a situation. Quantum energy fluxes have recently been computed numerically by \cite{Zilberman:2022a} in Kerr, who found that this is indeed the case. 
Another interesting question is how quantum matter would influence the angular momentum of spheres approaching the IH.

In this Letter, we compute the energy flux  $\langle \hat T_{vv} \rangle_{\rU}$  and angular momentum current-density $\langle \hat T_{v\varphi_-} \rangle_\rU$  (see Sec.~\ref{sec:geo} for the azimuthal coordinate $\varphi_-$ and  Eddington-type coordinate $v$)  components of the RSET for a real scalar quantum field in the Unruh state at the IH of KdS. 
The Unruh state is the relevant one since it models the late-time behaviour in gravitational collapse to a BH.
  
 Our numerical evidence suggests that
 $\langle \hat T_{V_-V_-} \rangle_{\rU} \sim V_-^{-2}$ as $V_-\to 0$, where $V_-\equiv -e^{-\kappa_-v}$  is a  Kruskal coordinate such that 
  $V_-=0$ on the IH and $\kappa_-$ is the surface gravity of the IH.
 For classical scalars on KdS, the divergence is weaker: $T_{V_-V_-} \sim V_-^{-(2-2\beta)}$,  where $\beta  \in (0,1/2)$ \cite{Hintz:2015,Dyatlov:2010, Hintz:2021,Dias:2018a}; similarly, changing the Unruh state to other initial Hadamard states would result in a correction of the same size as in the classical case \cite{KleinHintz}.
 We thus find the {\it leading} irregularity~\footnote{In a fully quantum description of both spacetime and matter, there would presumably be no divergence, i.e. we can only trust a semiclassical picture up to when the RSET assumes Planckian values.} of the RSET at the IH. 
We also find that $\langle \hat T_{V_-V_-} \rangle_{\rU}$ can
change sign with
 the polar angle $\theta$, differently from the classical case.
 
\begin{figure}
    \centering    \includegraphics[
    scale=0.3]{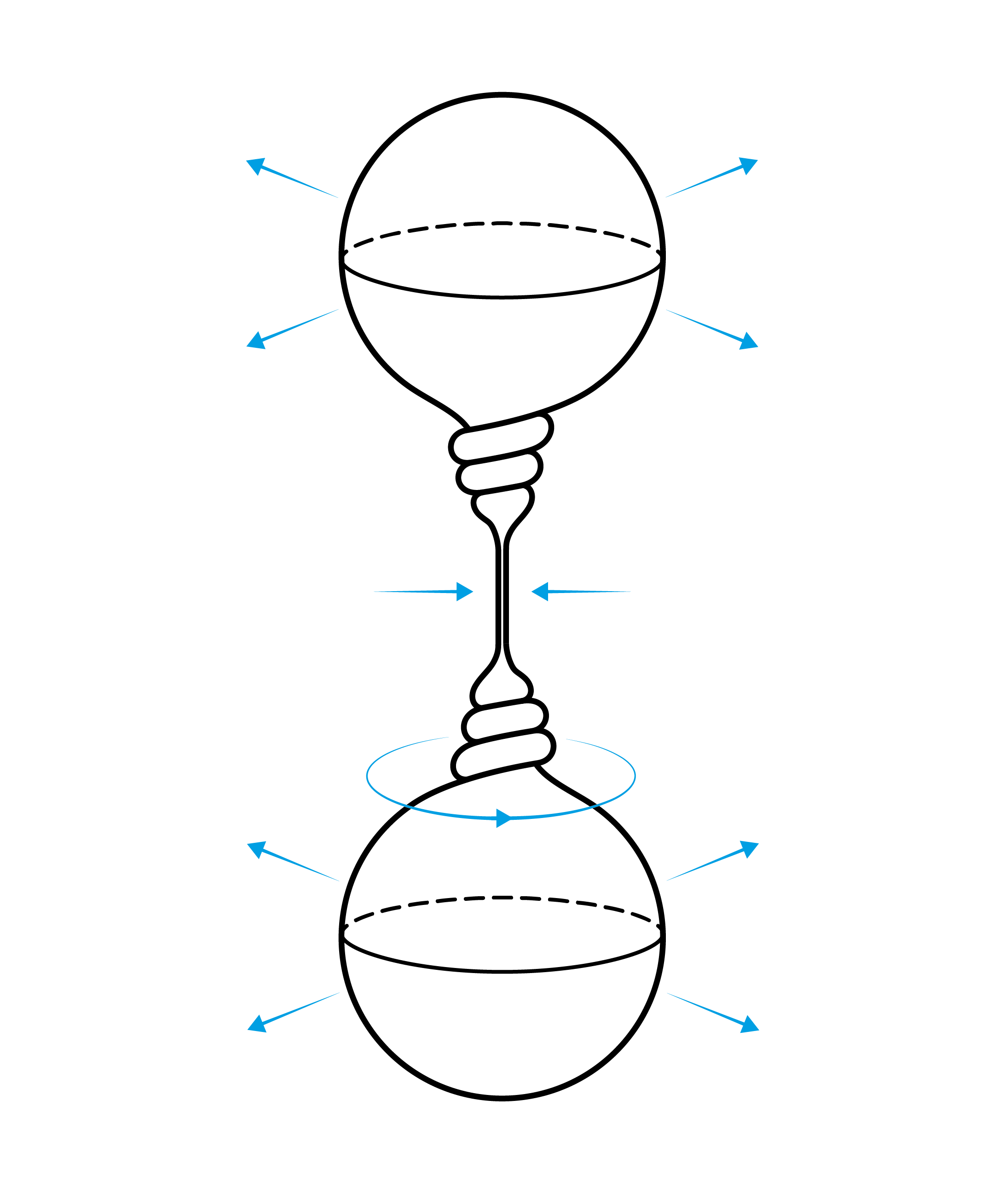}
    \caption{Infinite twisting between regions of diverging expansion and contraction on a cross section of the IH.}
    \label{fig:twisting}
\end{figure}

 In its turn, we find that the angular momentum current-density behaves as $\langle \hat T_{V_-\varphi_-} \rangle_\rU \sim V_-^{-1}$, which is still divergent at the IH but subdominant to $\langle \hat T_{V_-V_-} \rangle_{\rU}$.
  We find that either sign is possible depending on the values of the BH parameters and that, sufficiently close to extremality, the sign changes once between the poles and equatorial plane as the polar angle $\theta$ varies. 
  However, we also find
  that the angle average of $\langle \hat T_{V_-\varphi_-} \rangle_\rU$ near extremality is positive, 
  suggesting in view of standard flux-balance relations for the Komar 
  angular momentum that quantum effects tend to decrease the total angular momentum of $2$-spheres approaching the IH. Extrapolating 
  from our results on the fixed KdS background to a (hypothetical) solution
  of the semiclassical Einstein equations with backreaction, we find that the divergence of the angular momentum current-density will lead to diverging geometric twist of these $2$-spheres along latitudes separating infinite expansion and infinite contraction, see Fig.~\ref{fig:twisting}.

 The rest of this Letter is organized as follows.  Secs.~\ref{sec:geo} and~\ref{sec:field} respectively introduce KdS spacetime and scalar field quantum field theory on KdS. The
 formula for the RSET at the IH is 
 derived in Sec.~\ref{sec:SET}. Sec.~\ref{sec:num} contains our numerical results for the RSET. We summarize the results in Sec.~\ref{sec:sum}.
App.~\ref{sup} provides
 details on the background spacetime, the calculations of the RSET and the twisting of spheres.
 Throughout, we set $\hbar=c=G_N=1$.


\section{Geometric setup}
\label{sec:geo}

\begin{figure}
    \centering
    \includegraphics[width=0.4\textwidth]{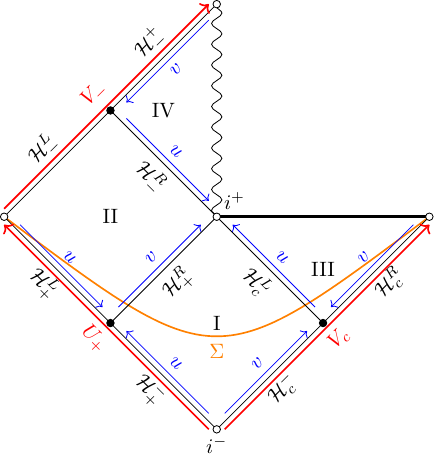}
    \caption{Carter-Penrose diagram of  KdS. The orange line $\Sigma$ is a Cauchy surface for regions $\rI$, $\rII$ and $\rIII$. The red lines indicate the ranges of different  Kruskal coordinates, while the blue lines illustrate $u$ and $v$ in the corresponding block. The wavy line represents the timelike ring singularity of KdS (at $r=0$ and $\theta=\pi/2$).}
    \label{fig:Penrose diag}
\end{figure}

In Boyer-Lindquist coordinates $(t,r,\theta,\varphi)$, the KdS metric describing a BH of mass $M$ with angular momentum parameter $a$ in the presence of a positive cosmological constant $\Lambda$ is
\begin{align}
    \label{eq:KdS metric}
     g&=\frac{\Delta_\theta a^2\sin^2\theta-\Delta_r}{\rho^2\chi^2}\td t^2 +\frac{\rho^2}{\Delta_r}\td r^2+\frac{\rho^2}{\Delta_\theta}\td \theta^2 \\\nonumber
     &+\left[\Delta_\theta(r^2+a^2)^2-\Delta_ra^2\sin^2\theta\right]\frac{\sin^2\theta}{\rho^2\chi^2}\td\varphi^2\\\nonumber 
     &+2\frac{a\sin^2\theta}{\rho^2\chi^2}[\Delta_r-\Delta_\theta(r^2+a^2)]\td t\td\varphi\, ,
\end{align}
where
\begin{align}	
\label{eq:D_r,D_t,rho,chi}
    \Delta_r&\equiv (1-\Lambda r^2/3)(r^2+a^2)-2Mr\, , 
   \quad
   \chi  \equiv 1+a^2\Lambda/3\, 
 ,\nonumber \\
     \Delta_\theta &\equiv 1 + (a^2 \Lambda/3) \cos^2\theta\,
 , \quad  
 \rho^2\equiv r^2+a^2\cos^2\theta\,
 .\nonumber
\end{align}
Henceforth, we set $M=1$ and restrict the parameters $\Lambda$ and $a$ to the sub-extremal range in which $\Delta_r(r)$ has three distinct positive roots
defining
the locations of the cosmological $(r_c)$,  event $(r_+)$ and inner $(r_-)$  horizons. The near-extremal regime is for $a$ close to its maximally allowed value, $a_{\text{max}}(\Lambda)$.
In that case, the Boyer-Lindquist coordinates cover the regions $\bR_t\times(r_i, r_j)\times\bS^2_{\theta,\varphi}$, with $r_i$ and $r_j$  two subsequent zeros of $\Delta_r$ (or $\pm\infty$).
Each horizon $r_j$, $j\in\{-,+,c\}$, has associated an angular velocity $\Omega_j$, a surface gravity $\kappa_j$ and
an azimuthal coordinate via $\td \varphi_j\equiv \td \varphi-\Omega_j\td t$. We also define a new radial coordinate via 
$\td r_*=\chi(r^2+a^2)\td r/\Delta_r$ and Eddington-type coordinates $v\equiv t+r_*$ and $u\equiv t-r_*$. Analytic continuation of the metric across $r=r_j$ is achieved in  Kruskal coordinates $(U_j, V_j)$~\cite{Borthwick:2018}, see App. \ref{sup}, from which the Carter-Penrose diagram is constructed -- see Fig. \ref{fig:Penrose diag}.
The region $\rI\cup\rII\cup\rIII$, together with the horizons $\mH_c^L$ and $\mH_+^R$ constitute a globally hyperbolic spacetime \cite{Klein:2022}: there exists a Cauchy surface (i.e., a smooth, spacelike hypersurface which is crossed exactly once by every inextendible causal curve) for this region.

We are interested in the angular momentum of the IH. Considering that, under backreaction, the BH becomes dynamic  and the spacetime is no longer a vacuum solution of the Einstein equation,  it seems reasonable to investigate a  quasilocal notion of angular momentum. Assuming that the spacetime remains axisymmetric, 
one plausible choice is the Komar angular momentum
\begin{align}
    J[{\mathcal S}] =&\frac{1}{16\pi}\int\limits_{{\mathcal S}} \nabla_\alpha  \psi_\beta \td\Sigma^{\alpha\beta}\, ,
\end{align}
of a topological sphere ${\mathcal S}$, where 
 $\psi^\mu \equiv \delta_\varphi^\mu$ is the rotational Killing field and $\td\Sigma^{\alpha\beta}$ an appropriately oriented covariant surface integration element. We choose orientations such that, in 
KdS, $J_{\text{KdS}}=\frac{a M}{\chi^2}$, which is independent of 
${\mathcal S}$. However, in a dynamical axisymmetric spacetime,
by Gauss' theorem and C.3.6 of \cite{Wald:1984}
\begin{align}
\label{eq:flux}
    J[{\mathcal S}_2]-J[{\mathcal S}_1] 
    = -\int\limits_{\Delta} (T_{\alpha\beta} -\tfrac{1}{2}g_{\alpha\beta} T^\sigma{}_\sigma-\Lambda g_{\alpha\beta})  \psi^\alpha \td\Sigma^\beta.
    \end{align}
    Here, $\Delta$ is a 3-dimensional surface tangent to $\psi^\alpha$
bounding ${\mathcal S}_1$ and ${\mathcal S}_2$, and 
$\td \Sigma^\alpha$ is an appropriately oriented covariant integration element on $\Delta$. We are interested in the case where 
${\mathcal S}_2$
approaches the future IH, ${\mathcal H}^R_-$, at constant value of $u$, while ${\mathcal S}_1$, located at the same constant $u$, stays at some constant value $V_-<0$. We will assume 
that, to estimate quantum effects, the stress tensor can be replaced in the small backreaction regime by the RSET $\VEV{\hat T_{\alpha\beta}}_\rU$; we denote its value at the IH by $\VEV{\hat T_{\alpha\beta}(\theta)}^{\rm IH}_{\rU}$.

To evaluate Eq.~\eqref{eq:flux}, we require the RSET component $\VEV{\hat T_{\varphi_- v}}_{\rU}$. We present evidence in the following sections that this quantity approaches a finite value at the IH.
Using this result and  Eq.~\eqref{eq:KdS metric}, we obtain
\begin{equation}
\label{JSv}
 J[{\mathcal S}(u,v)] \approx - v \langle \langle \hat T_{v\vp_-} \rangle \rangle^{\rm IH}_{\rU} 
\end{equation}
as $v \to \infty$ at constant 
$u$, where ${\mathcal S}(u,v)$ is a 2-sphere of constant values $u,v$. The angle-averaged expectation value of the RSET  at the IH  in Eq.~\eqref{JSv} is defined by~\footnote{Note that $\langle\langle \hat T_{v\vp_-} \rangle\rangle^{\rm IH}_{\rU}$
is independent of the Killing parameter $u$ because the Unruh state is stationary.}
\begin{align}
\label{eq:integral over theta and phi}
   \langle\langle \hat T_{v\vp_-} \rangle\rangle^{\rm IH}_{\rU}\equiv 
   2\pi\frac{r_-^2+a^2}{\chi} \int\limits_0^\pi\td \theta \sin\theta \VEV{\hat T_{v\vp_-}(\theta)}^{\rm IH}_{\rU}\, .
\end{align}


\section{The scalar quantum field}
\label{sec:field}

Consider a minimally coupled scalar field $\phi$ of mass $\mu=\sqrt{2\Lambda/3}$, which satisfies the same equations of motion as the massless, conformally coupled scalar field,
\begin{align}
    \label{eq: KG eqn}
    (g^{\alpha\beta}\nabla_\alpha \nabla_\beta-\mu^2)\phi=0\, .
\end{align}
On a single Boyer-Lindquist block, this differential equation reduces to a set of ordinary differential equations via
\begin{align}
\label{eq:modeansatz}
  \phi_{\omega\ell m}(t,r,\theta,\vp)={\mathcal N}_{\omega\ell m}
  e^{-i\omega t}e^{i m\vp}S_{\ell m}^\omega(\theta)R_{\omega\ell m}(r)\, .
\end{align}
Here, $m\in \mathbb{Z}$, $\ell\in\mathbb{N}_{\geq |m|}$,  $S_{\ell m}^\omega(\theta)$ are the spheroidal wave functions~\cite{Suzuki:1998}
and $R_{\omega\ell m}(r)$ obey a Schr\"odinger-like equation~\cite{Suzuki:1998, Suzuki:1999}.

Consider a complete set  $\{\phi_I\}_I$, of  mode solutions to the Klein-Gordon equation \eqref{eq: KG eqn} 
which is orthonormal with respect to the usual Klein-Gordon inner product. 
Then one can quantize the  field by expanding it as
\begin{align}\label{eq:quantum field}
    \hat \phi(x)=\sum\limits_I \left(\phi_I(x) \hat b_I+\overline{\phi_I}(x)\hat b^\dagger_I\right)\, .
\end{align}
The coefficients $\hat{b}_I$ and $\hat{b}^\dagger_I$ are creation and annihilation operators, and the vacuum state $|{\rm U}\rangle$ of the Fock space on which these operators act satisfies $\hat b_I|{\rm U}\rangle=0$ for all $I$.

The Unruh state can be constructed in this way by choosing $\phi_I$ to be the Unruh modes: (i) modes which are positive frequency with respect to $U_+$ on $\mH_+^L \cup \mH_+^-$ and  vanish on $\mH_c^-\cup\mH_c^R$;
  together with (ii) modes  which are positive frequency with respect to $V_c$ on $\mH_c^- \cup \mH^R_c$ and vanish on $\mH_+^-\cup\mH_+^L$.

However, the Boulware modes $\phi^{\ri/\ru,\rI}_{\omega\ell m}$ (with reflection and transmission coefficients $\mathcal{R}^{\ri/\ru,\rI}_{\omega\ell m}$ and $\mathcal{T}^{\ri/\ru,\rI}_{\omega\ell m}$) and $\phi^{\ri/\ru,\rII}_{\omega\ell m}$, defined in App.~\ref{sup}, are easier to calculate than the Unruh modes.
By expressing the Unruh modes in terms of Boulware modes similarly as in~\cite{Klein:2021, Zilberman:2022}, we obtain for the anti-commutator in  $\rII$:
\begin{align}\label{eq:twopt}
    &\VEV{\hat \phi(x) \hat \phi(y)}_\rU^{(s)} \equiv 
   \frac{1}{2}\VEV{\left(\hat \phi(x)\hat \phi(y)+\hat \phi(y)\hat \phi(x)\right)}_\rU
    =
 \\   \nonumber
  &  \frac{1}{2}\sum\limits_{\ell, m}\int\limits_0^\infty \td \omega
    \left[\coth\left(\tfrac{\pi\omega_c}{\kappa_c}\right)\abs{\mathcal{T}^{\ri,\rI}_{\omega\ell m}}^2\abs{\frac{\omega_+}{\omega_c}}\left\{\phi^{\ri,\rII}_{\omega\ell m}(x),\overline{\phi^{\ri,\rII}_{\omega\ell m}}(y)\right\}\right.\\\nonumber
   &+\coth\left(\tfrac{\pi\omega_+}{\kappa_+}\right)\left[\left\{\phi^{\ru,\rII}_{\omega\ell m}(x), \overline{\phi^{\ru,\rII}_{\omega\ell m}}(y)\right\}\right.\\\nonumber
   &\left.+\abs{\mathcal{R}^{\ru,\rI}_{\omega\ell m}}^2\left\{\phi^{\ri,\rII}_{\omega\ell m}(x),\overline{\phi^{\ri,\rII}_{\omega\ell m}}(y)\right\}\right] \\\nonumber
   &+\left. 2\csch\left(\tfrac{\pi\omega_+}{\kappa_+}\right)\Re\left[\mathcal{R}^{\ru,\rI}_{\omega\ell m}\left\{\phi^{\ri,\rII}_{\omega\ell m}(x),\overline{\phi^{\ru,\rII}_{\omega\ell m}}(y)\right\}\right]\right],
\end{align}
where $\{f(x),g(y)\}\equiv f(x)g(y)+f(y)g(x)$ for two functions $f$ and $g$ and $\omega_j(\omega,m)\equiv \omega-m\Omega_j$ for 
$j=-,+,c$. 

\section{The stress-energy tensor}
\label{sec:SET}

The quantum observable corresponding to the classical stress-energy tensor   in some quantum state $\Psi$ is the RSET $\langle \hat T_{\mu\nu} \rangle_{\Psi}$. 
We wish to see if, and how, the RSET in the Unruh state ($|\Psi\rangle = |{\rm U}\rangle$) diverges. 

At least for sufficiently small $a$ or $\Lambda$, the Unruh state is Hadamard up to but not including the IH~\cite{Klein:2022}.
To understand the divergent behavior of $\langle \hat T_{\mu\nu} \rangle_{\rm U}^{\rm IH}$, it 
suffices to calculate the offset to the expectation value, 
$\langle \hat T_{\mu\nu} \rangle_{\rm C}$, in 
some comparison state $|{\rm C}\rangle$ which is Hadamard  in an open two-sided neighborhood of the IH. Then -- see, e.g., \cite{Hollands:2001nf} -- $\langle \hat T_{\mu\nu} \rangle_{\rm C}$ is smooth in a two-sided neighborhood of the IH in Kruskal coordinates, and so we must have $\langle \hat T_{vz} \rangle_{\rm C}^{\rm IH}=0$, for $z\in\{v,\vp_-\}$. Thus,
\begin{align}
\label{eq:ptsplit}
    &\VEV{\hat T_{vz}(x)}_{\rm U}^{\rm IH}
    = \\ & 
     \lim_{x\to \mH_-^{L,R}} 
    \lim_{x'\to x} D_{vz'}
    \left(
\VEV{\hat \phi(x)\hat \phi(x')}_{\rm U}^{(s)}-
\VEV{\hat \phi(x)\hat \phi(x')}_{\rm C}^{(s)}
    \right),\nonumber
\end{align}
where $D_{\alpha \beta'}$ is a bi-differential operator given in App.~\ref{sup} and $x$ is any point where {\it both} states are defined.

We construct the comparison state similarly to \cite{Hollands:2019} for RN: we modify the metric in $r<\delta<r_-$, with $0<\delta\ll r_-$, 
replacing the singularity at $r=0$.
Combined with Eqs.~\eqref{eq:twopt} and \eqref{eq:ptsplit}, this yields (App.~\ref{sup}):
\begin{align}
\label{eq: formula SET}
    &\VEV{\hat T_{vz}(\theta)}_\rU^{\rm IH}=\frac{\chi}{4\pi (r_-^2+a^2)}\left\{\sum\limits_{\ell=0}^\infty\int\limits_0^\infty \frac{\td \omega_-}{\omega_-}F^{z}_{\ell\, 0}(\omega_-,\theta)\right.\nonumber \\
    &\left.+\sum\limits_{\ell\geq m>0}\int\limits_0^\infty \frac{\td \omega_-}{\omega_-}\left(F^z_{\ell m}(\omega_-,\theta)-F^z_{\ell m}(-\omega_-,\theta)\right)\right\}\, ,
    \end{align}
   \begin{align}
    &F_{\ell m}^z(\omega_-,\theta)\equiv \omega_-^2\frac{c_z}{\omega_+}\abs{S^\omega_{\ell m}(\theta)}^2
   \left[ 
   \frac{-\omega_+}{\omega_-}\coth\left(\tfrac{\pi\omega_-}{\kappa_-}\right)\right.
   \nonumber
    \\\nonumber
    &
    +\coth\left(\tfrac{\pi\omega_+}{\kappa_+}\right)\left(\abs{\mathcal{R}^{\ri,\rII}_{\omega\ell m}}^2+\abs{\mathcal{R}^{\ru,\rI}_{\omega\ell m}}^2 \abs{\mathcal{T}^{\ri,\rII}_{\omega\ell m}}^2\right)
    \\\nonumber
    &+\left.2\csch\left(\tfrac{\pi\omega_+}{\kappa_+}\right)\Re\left[\mathcal{R}^{\ru,\rI}_{\omega\ell m}\mathcal{R}^{\ri,\rII}_{\omega\ell m}\mathcal{T}^{\ri,\rII}_{\omega\ell m}\right]\right.\\
    &+\left. \coth\left(\tfrac{\pi\omega_c}{\kappa_c}\right)\left(1-\abs{\mathcal{R}^{\ru,\rI}_{\omega\ell m}}^2\right)\abs{\mathcal{T}^{\ri,\rII}_{\omega\ell m}}^2
    \right],
         \label{eq:SET integrand}
\end{align}
where $z\in\{v,\vp_-\}$, $c_v\equiv \omega_-$ and $c_{\vp_-}\equiv -m$.

\section{Numerical results}
\label{sec:num}

\begin{figure}
    \centering
    \includegraphics[width=0.47\textwidth]{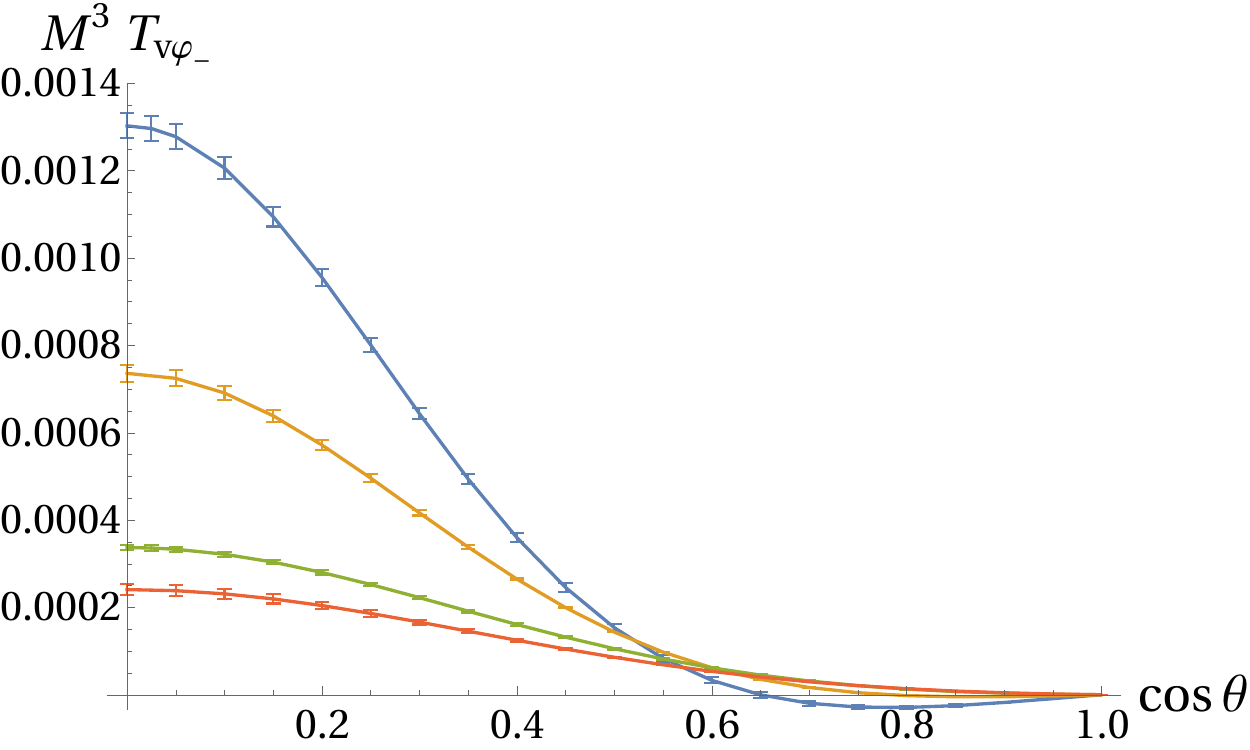}
    \caption{The $\cos\theta$-dependence of $\VEV{T_{v\vp_-}}_\rU^{\rm IH}$ for  $\Lambda =1/30$ and $a=0.95$  (blue), $0.975$ (orange), $1$ (green) and $1.007$  (red). Note that $a_{\text{max}}\approx1.012$.}
    \label{fig:Tvp tdep}
\end{figure}

We plot in Fig.~\ref{fig:Tvp tdep} the angular momentum current at the IH as a function of $\cos\theta$ for $\Lambda=1/30$ and $a=0.95$, $0.975$, $1$, and $a_{\text{max}}-1/200$, where $a_{\text{max}}\approx 1.012$. 
In App.~\ref{sup}, we also plot the angular momentum current  at $\cos\theta=0.9$ as a function of $a$ for different values of  $\Lambda$.
Our first main result is that $\VEV{T_{v\vp_-}}_\rU^{\rm IH}$ can have different signs depending on the latitude $\theta$. 

The integral  \eqref{eq:integral over theta and phi}
 determines the change of the  quasilocal angular momentum near the IH (Sec.~\ref{sec:geo}). Our second main result is that, for all near-extremal values of $a$  in Fig.~\ref{fig:Tvp tdep}, we find that this integral is positive: $\langle\langle \hat T_{v\vp_-} \rangle\rangle^{\rm IH}_{\rU}=0.00667\pm 0.00017$, $0.004769\pm0.000074$, $0.002974\pm 0.000048$ and $0.002398\pm 0.000039$ for, respectively, $a=0.95$, $0.975$, $1$ and $a_{\text{max}}- 0.005$.
Substituting these positive values into Eq.~\eqref{JSv}, we conclude that the angular momentum $J[{\mathcal S}(u,v)]$ of the quantum field, associated with a sphere of constant $v$ and  $u$, is going to $-\infty$ linearly in $v$ as the IH is approached, $v \to \infty$. Thus, it is of the {\it opposite} sign as the angular momentum of such a sphere in the background ($\equiv Ma/\chi^2$).

\begin{figure}
    \centering
    \includegraphics[width=\linewidth]{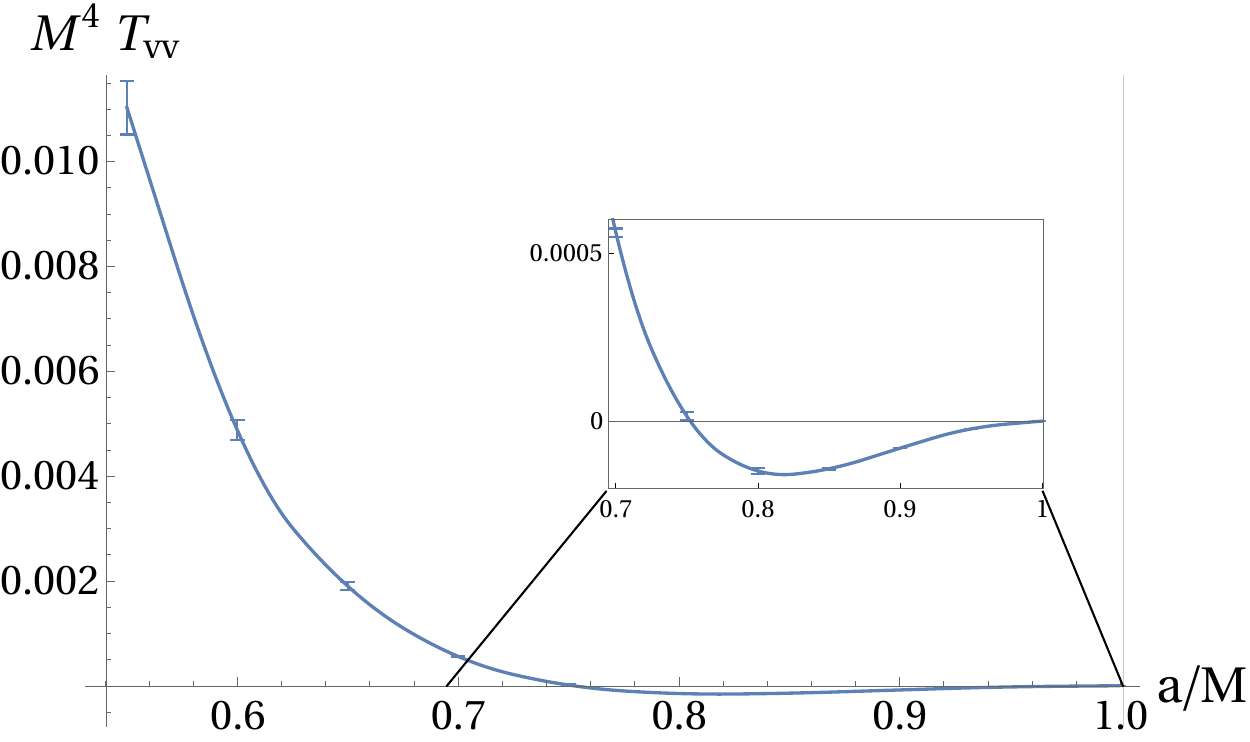}
    \caption{$\VEV{\hat T_{vv}}_{\rU}^{\rm IH}$ for $\theta=0$, $\Lambda=\frac{1}{270}$ as a function of  $a$. The vertical line indicates $a_{\text{max}}$. The inset is a closeup  near extremality.}
    \label{fig: Tvv axis}
\end{figure}

In Fig.~\ref{fig: Tvv axis} we plot $\VEV{\hat T_{vv}}_{\rU}^{\rm IH}$ on the rotation axis as a function of $a$.
To achieve better comparability with the corresponding results obtained in RN(-dS)~\cite{Zilberman:2019, Hollands:2020} and Kerr \cite{Zilberman:2022a} for a minimally-coupled scalar field of mass $\mu^2=2\Lambda/3$ (which vanishes in RN and Kerr), we have set $\Lambda=1/270$. We see qualitatively very similar results. In particular, 
$\VEV{\hat T_{vv}}_{\rU}^{\rm IH}$ begins positive for small 
$a$, changes sign at an intermediate value, and then approaches zero from below as $a$ approaches $a_{\text{max}}$.  For the same parameter values as the orange curve in Fig. \ref{fig:Tvp tdep}, we have also checked that $\VEV{\hat T_{vv}}_{\rU}^{\rm IH}$
changes sign with $\theta$, qualitatively similar to \cite{Zilberman:2022a}.

\section{Conclusions}
\label{sec:sum}

We have computed the $vv$- and $v\vp_-$-components of the RSET in the Unruh state on the IH of a KdS BH. These components constitute, respectively, the leading divergence of the $V_-V_-$ and $V_-\varphi_-$ components of the RSET on the IH under a tensorial transformation with $\partial V_-/\partial v \sim V_-$.

The divergence of $\VEV{\hat T_{V_-V_-}}_\rU$ on the IH ($V_-=0$) is proportional to $V_-^{-2}$ with a proportionality factor that is generically nonzero, at least on the axis of rotation. The sign of this factor can change with  $a$ and $\theta$, similarly as in Kerr \cite{Zilberman:2022a}.

The divergence of $\VEV{\hat T_{V_-\varphi_-}}_\rU$ on the IH is proportional to $V_-^{-1}$, again with a
generically nonzero  proportionality factor. Interestingly, the sign of this quantity can change with the latitude on the sphere, $\theta$, near extremality. 
 Nonetheless, we find that the {\it angle average} of $\VEV{\hat T_{V_-\varphi_-}}_\rU$ diverges near extremality to $-\infty$ for our chosen parameters. This indicates that, if backreaction effects were taken into account, the {\it total} angular momentum of such a sphere would suffer an $O(1)$ decrease relative to the background value $J_{\rm KdS}=Ma/\chi^2$ in our conventions where $a>0$, see Eq. \eqref{JSv}, by which time the background underlying our calculation would have to be updated. This
 resembles the behavior of the total scalar charge of a charged quantum scalar field near the IH of a near-extremal RN BH~\cite{Klein:2021a}, and indicates the absence of `runaway' behavior for the charge and angular momentum. 

Combined with the results of \cite{KleinHintz}, the leading divergence of the RSET that we find is  
not special for the Unruh state but true for an {\it arbitrary} initial (Hadamard) state, and stronger than the divergence of the classical stress-energy, at least in the parameter regime where KdS is mode-stable  
\footnote{For {\it partial}  mode stability results of KdS, see~\cite{2022CMaPh.394..797C}. }.

As we argue in App.~\ref{sup}, if we extrapolate the qualitative features of our findings to a setting with backreaction, then topological spheres approaching the IH undergo infinite expansion in some parts, and contraction in others. On a latitude separating such regions, we have a diverging relative twisting of neighboring latitudes and/or a twisting of perpendicular incident lightrays towards or against the $\varphi_-$-direction, Fig.~\ref{fig:twisting}.

\section*{Acknowledgements}
C.K., M.C. and S.H. would like to thank, respectively, B.~Bonga, A.~Ori and  J.~Zahn for helpful discussions.
We thank Thomas Endler for creating Fig.~\ref{fig:twisting} based on our suggestions and for giving us permission to include it in this publication.
C.K. and S.H. thank the Erwin Schrödinger Institut in Vienna, where part of this work has been completed, for its hospitality and support.
S.H. is grateful to the Max-Planck Society for supporting the collaboration between MPI-MiS and Leipzig U., grant Proj. Bez. M.FE.A.MATN0003
This work has been funded by the Deutsche Forschungsgemeinschaft (DFG) under the Grant No. 406116891 within the Research Training Group RTG 2522/1.

\newpage
\clearpage
\onecolumngrid
\appendix

\section{Supplemental Material}\label{sup}

In this supplemental material we give a more detailed account of some of the calculations and definitions described in the main paper, and we present more details of our numerical analysis. In Sec.~\ref{sec: add geo}, we present  additional information on the background spacetime and the Kruskal coordinates. Sec.~\ref{sec: add EoM} contains a detailed description of the angular and radial Teukolsky equations and their mode solutions. The derivation of the quantum RSET is described in Sec.~\ref{sec:RSET}. Finally, the details of the numerical implementation are described in Sec.~\ref{sec: add num}.
We set $M=1$ when giving parameter values as in the main document.

\subsection{Additional information on the geometric setup}
\label{sec: add geo}

In this section, we give additional details and definitions for the geometric setup, in particular the Kruskal-type coordinates for Kerr-de Sitter spacetime. A more detailed discussion can also be found in \cite{Borthwick:2018}.

To begin, the angular velocity of the horizon $\mH_j$, $j\in\{-,+,c\}$ mentioned in the main text is given by
\begin{align}
    \Omega_j\equiv \frac{a}{r_j^2+a^2}\, ,
\end{align}
and we have defined the surface gravities of the horizons as \footnote{The choice of surface gravity and angular velocity depends on the choice of normalization of the Killing vector field generating the horizon. In the literature, there exist different choices in particular for the distribution of factors of $\chi$, compare for example \cite{Borthwick:2018} and \cite{Gregory:2021}.}
\begin{align}
    \kappa_j\equiv \frac{\abs{\del_r\Delta_r}_{r=r_j}}{2\chi(r_j^2+a^2)}\, .
\end{align}
With the help of these quantities, we can now define a set of Kruskal coordinates $\{U_j,V_j\}$ for each  horizon $\mH_j$, $j\in\{-,+,c\}$. 

In the two most relevant blocks $\rI=\{r_+<r<r_c\}$,  with the time orientation chosen such that $t$ is increasing towards the future, and $\rII=\{r_-<r<r_+\}$ with $r$ decreasing towards the future, the Kruskal coordinates are related to the Eddington-type coordinates $u$ and $v$ defined in the main text by
\begin{subequations}
\begin{align}
    U_+&=-e^{-\kappa_+ u}\,, &  V_+&=e^{\kappa_+ v}\, ,\\
    U_c&=e^{\kappa_c u}\, , & V_c&=-e^{-\kappa_c v}\, ,
\end{align}
\end{subequations}
in $\rI$, and 
\begin{subequations}
\begin{align}
 \label{eq:K2E}
    U_+&=e^{-\kappa_+u}\, , & V_+&=e^{\kappa_+ v}\, ,\\
    \label{eq:K2E1}
    U_-&=-e^{\kappa_- u}\, , & V_-&=-e^{-\kappa_- v}\, , 
    \end{align}
\end{subequations}
in $\rII$. Of particular relevance here is $V_-$, which vanishes on the inner horizon (IH) $\mH_-^R$.

With the help of the coordinate systems $(U_j, V_j, \theta, \varphi_j)$, the metric can be analytically extended across the bifurcate Killing horizon at $r=r_j$, including the bifurcation sphere $\{U_j=V_j=0\}$ \cite{Borthwick:2018}. These Kruskal-type blocks can be glued together to form part of the maximal analytic extension. The Carter-Penrose diagram of the (physical) part of this extension considered in the main paper is shown in its Fig.~\ref{fig:Penrose diag}.

For completeness, let us also give the metric in the coordinates $(u,v,\theta, \vp_j)$, which are used for most of the computations. In these coordinates, the metric takes the form
\begin{align}
     g=&\frac{ \Delta_\theta \sin^2\theta a^2 \left( r_j^2-r^2\right)^2 - \Delta_r  \rho_j^4}{4\chi^2\rho^2(r_j^2+a^2)^2}  (\td u + \td v)^2+\frac{\Delta_r\rho^2}{4\chi^2(r^2+a^2)^2}(\td u-\td v)^2+ \frac{\rho^2}{\Delta_\theta} \td \theta^2 + g_{\varphi\varphi}\td \varphi_j^2 \\\nonumber
    &+ \frac{a\sin^2\theta\left[\Delta_r \rho_j^2 - (r^2+a^2)\Delta_\theta \left(r_j^2-r^2 \right) \right]}{\chi^2\rho^2(r_j^2+a^2)} \td \varphi_j (\td u + \td v)\, ,
\end{align}
where $\rho_j^2\equiv \rho^2(r_j,\theta)$. Note that the metric components $g_{uu}$ and $g_{vv}$ vanish like $(r-r_j)^2$ as $r\to r_j$, while $g_{uv}$, $g_{u\vp_j}$, and $g_{v\vp_j}$ vanish like $(r-r_j)$.

\subsection{Mode solutions to the Teukolsky equation}
\label{sec: add EoM}
 In this section, we present a more detailed description of the mode solutions to the Klein-Gordon equation on Kerr-de Sitter.
 As explained in the main text, using an ansatz of the form 
 \begin{align}
 \label{eq:modeansatz}
 \phi_{\omega\ell m}(t,r,\theta,\vp)=\mathcal{N} e^{-i\omega t}e^{i m\vp}S_{\ell m}^\omega(\theta)R_{\omega\ell m}(r)\, 
 \end{align}
 where $\mathcal{N}$ is a normalization constant,
 the Klein-Gordon equation (Eq.~\eqref{eq: KG eqn} in the main text) separates into the angular Teukolsky equation \cite{Suzuki:1998}
\begin{align}
\label{eq: angular TE}
   &\left[ \del_x\left(\Delta_\theta(1-x^2)\del_x\right)+E_{\ell m}^\omega-2a^2\lambda x^2-\frac{\chi^2(a\omega+m)^2}{\Delta_\theta}\right.\\\nonumber
   &\left.+\frac{\chi^2x^2}{\Delta_\theta}\left((a\omega )^2-\frac{m^2}{1-x^2}\right)\right]S^\omega_{\ell m}(x)=0\, ,
\end{align}
with $x\equiv \cos\theta$, and the radial Teukolsky equation \cite{Suzuki:1998, Suzuki:1999}
\begin{align}
\label{eq: radial TE}
   & \left[\del_r\Delta_r\del_r+\frac{\chi^2K^2}{\Delta_r}-E^\omega_{\ell m}-\mu^2 r^2\right]R_{\omega\ell m}(r)=0\, ,\\\nonumber
   & K\equiv \omega(r^2+a^2)-am\, .
\end{align}

The two equations are connected by the angular eigenvalues $E_{\ell m}^\omega$ which are determined by demanding that the solution to the angular equation is regular at both $\theta=0$ and $\theta=\pi$. The corresponding solutions to the angular Teukolsky equation are called spheroidal wave functions. 
We normalize the spheroidal wave functions so that
\begin{align}\label{eq:norm S}
    \int\limits_0^\pi \td \theta \sin\theta \overline{S^\omega_{\ell m}}(\theta)S^\omega_{\ell^\prime m^\prime}(\theta)=\delta_{\ell\ell^\prime}\delta_{mm^\prime}\, ,
\end{align}
and note that they satisfy $S^\omega_{\ell m}(\theta)=(-1)^{(\ell+m)}S^{\omega}_{\ell m}(\pi-\theta)$.

The solutions to the radial equation can, for example, be specified by their asymptotic behaviour at the various horizons. One particular set of such solutions in region $\rI$ are the Boulware modes used for computational purposes in the main document. They are defined by their asymptotic behavior at $\mH_c^-\cup\mH_+^-$ or, in other words, by fixing the asymptotic behaviour of the radial function $R_{\omega\ell m}(r)$ for $r\to r_+$ and $r\to r_c$. In particular, there are two types of these modes: the "up"- modes, which behave like 
\begin{align}
    R^{\ru,\rI}_{\omega\ell m}\sim \begin{cases} (r_+^2+a^2)^{-1/2}\left(e^{i\omega_+ r_*}+\mathcal{R}^{\ru,\rI}_{\omega\ell m}e^{-i\omega_+ r_*}\right) & r\to r_+ \\ (r_c^2+a^2)^{-1/2} \mathcal{T}^{\ru,\rI}_{\omega\ell m} e^{i\omega_c r_*} & r\to r_c\end{cases}
\end{align}
 and the "in"-modes, which behave like
\begin{align}
    R^{\ri,\rI}_{\omega\ell m}\sim \begin{cases} (r_+^2+a^2)^{-1/2}\mathcal{T}^{\ri,\rI}_{\omega\ell m} e^{-i\omega_+ r_*} & r\to r_+\\ (r_c^2+a^2)^{-1/2}\left( e^{-i\omega_c r_*}+\mathcal{R}^{\ri,\rI}_{\omega\ell m} e^{i\omega_c r_*}\right) & r\to r_c\, .\end{cases}
\end{align}
The coefficients $\mathcal{R}^{\ri/\ru,\rI}_{\omega\ell m}$ and $\mathcal{T}^{\ri/\ru,\rI}_{\omega\ell m}$ are, respectively, the reflection and transmission coefficients in region $\rI$. 

A separate set of Boulware modes can also be defined in the black hole interior region $\rII$. In a similar way as before, they are fixed by requiring
\begin{align}\label{eq:asympt Rin}
 R^{\ri,\rII}_{\omega\ell m}\sim \begin{cases}
     (r_+^2+a^2)^{-1/2}e^{-i\omega_+ r_*} & r\to r_+ \\
     (r_-^2+a^2)^{-1/2}\left(\mathcal{R}_{\omega\ell m}^{\ri,\rII}e^{i\omega_-r_*}+\mathcal{T}^{\ri,\rII}_{\omega\ell m}e^{-i\omega_- r_*}\right) & r\to r_-
 \end{cases}   
\end{align}
or
\begin{align}
  R^{\ru,\rII}_{\omega\ell m}\sim \begin{cases}
     (r_+^2+a^2)^{-1/2}e^{i\omega_+ r_*} & r \to r_+ \\
     (r_-^2+a^2)^{-1/2}\left(\mathcal{T}_{\omega\ell m}^{\ru,\rII} e^{i\omega_-r_*} +\mathcal{R}^{\ru,\rII}_{\omega\ell m} e^{-i\omega_- r_*} \right) & r\to r_-\, . \end{cases}  
\end{align}

The scattering coefficients can be related to each other by considering the Wronskian $W[\tilde{R}_1,\tilde{R}_2]=\tilde{R}_1\del_{r_*}\tilde{R}_2-\tilde{R}_2\del_{r_*}\tilde{R}_1$ of two radial solutions $R_1$ and $R_2$, where, $\tilde{R}_{1,2}\equiv (r^2+a^2)^{1/2} R_{1,2}$.
This Wronskian is independent of $r_*$. Comparing the results in the two asymptotic regimes, one finds, suppressing the indices $\omega$, $\ell$ and $m$ for brevity,
\begin{align}
\label{eq: Wronskian ext}
    \abs{\mathcal{R}^{\ri,\rI}}^2+\frac{\omega_+}{\omega_c}\abs{\mathcal{T}^{\ri,\rI}}^2=\abs{\mathcal{R}^{\ru,\rI}}^2+\frac{\omega_c}{\omega_+}\abs{\mathcal{T}^{\ru,\rI}}^2=1\, ,\\
    \label{eq: Wronskian int}
    \abs{\mathcal{T}^{\ri/\ru,\rII}}^2-\abs{\mathcal{R}^{\ri/\ru,\rII}}^2=\frac{\omega_+}{\omega_-}\, .
\end{align}

Finally, one can choose the normalization constant by demanding that the modes are normalized with respect to the usual Klein-Gordon inner product,
\begin{align}
  -i\,\sigma\left(\phi^{\ri/\ru,\rI}_{\omega\ell m},\overline{\phi^{\ri/\ru,\rI}_{\omega^\prime\ell^\prime m^\prime}}\right)=\delta_{\ell\ell^\prime}\delta_{mm^\prime}\delta(\omega-\omega^\prime)\text{sign}(\omega_{c/+}) \, . 
\end{align}
Here, the symplectic form $\sigma$ of a pair $\psi,\phi$ of  solutions to the Klein-Gordon equation
\begin{align}
    \sigma(\phi, \psi)=\int\limits_\Sigma (\phi \nabla_\alpha \psi-\psi\nabla_\alpha \phi)\td \Sigma^\alpha \, ,
\end{align}
 is defined on the Cauchy surface $\Sigma$ -- in this case for region $\rI$ -- with future-pointing covariant surface integration element $\td\Sigma^\alpha$. By a limit deformation, $\Sigma$ may be taken to be $\mH_+^-\cup \mH_c^-$.

 This normalization condition is satisfied by setting the normalization constant to be $\mathcal{N}^{\ri,\rI}_{\omega\ell m}=\sqrt{\frac{\chi}{4\pi\abs{\omega_c}}}$ and $\mathcal{N}^{\ru,\rI}_{\omega\ell m}=\sqrt{\frac{\chi}{4\pi\abs{\omega_+}}}$ for the Boulware modes in region $\rI$, and  $\mathcal{N}^{\ri/\ru,\rII}_{\omega\ell m}=\sqrt{\frac{\chi}{4\pi\abs{\omega_+}}}$ for the Boulware modes in region $\rII$.

Using the symmetry properties of the radial and angular equations and the fact that the chosen normalization constant $\mathcal{N}_{\omega\ell m}^{\ri/\ru,\rI/\rII}$ is symmetric under the simultaneous sign flip $(\omega,m)\to-(\omega,m)$, we obtain that the Boulware modes thus constructed satisfy $\phi^{\ri/\ru,\rI}_{(-\omega) \ell (-m)}=(-1)^m\overline{\phi^{\ri/\ru,\rI}_{\omega\ell m}}$. The factor $(-1)^m$ arises from the spheroidal wave function.


\subsection{The derivation of the stress-energy tensor}
\label{sec:RSET}

Classically, the stress-energy tensor for a minimally coupled scalar field of mass $\mu$ is given by
\begin{align}
\label{eq:Tdef}
T_{\alpha\beta}=\del_\alpha\phi(x)\del_\beta\phi(x)-\frac{1}{2}g_{\alpha\beta}\left(\del_\nu\phi(x)\del^\nu\phi(x)+\mu^2\phi^2(x)\right)\, .
\end{align}

As a result, the expectation value of the corresponding quantum observable can be written as in Eq.~\eqref{eq:ptsplit} of the main document, where the bi-differential operator $D_{\alpha \beta'}(x,x')$ for a general coordinate component and spacetime point is given by 
\begin{align}
\label{eq:Ddef}
    D_{\alpha \beta'}(x,x')=& \partial_\alpha \partial_{\beta'}-\frac{1}{2}g_{\alpha\beta'}(x,x')\left(g^{\gamma \delta'}(x,x') \partial_\gamma \partial_{\delta'}+\mu^2\right)\, ,
\end{align}
where $g_\nu^{\mu^\prime}(x,x')$ is the bivector of parallel transport mapping (co-)vectors at $x'$ to (co-)vectors at $x$.
If we consider only points on the IH and restrict to the $vz$-component of the stress-energy tensor, with $z\in \{v,\vp_-\}$, then $D_{\alpha\beta'}$ can be simplified to
\begin{align}
\label{eq:Dsimple}
    D_{vz'}=\partial_v\partial_{z'}-\frac{\delta_{z\vp_-}}{2}g_{v\vp_-}g^{uv}\partial_u\partial_{v'}\, .
\end{align}

As described in Eq.~\eqref{eq:ptsplit} of the main text, we compute the RSET at the IH in the Unruh state using state subtraction. Since the comparison state used in the state subtraction is chosen to be Hadamard in a two-sided neighbourhood of the horizon, the expectation value of the RSET in this state must be regular across the horizon, implying that the expectation value of the $vv$- and $v\vp_-$components vanish. Therefore, this method allows us to extract the expectation values of these components in the Unruh state at the IH.  As described in the main document, the comparison state is constructed on a modified spacetime by expanding the quantum field in terms of modes,
\begin{align}
    \hat \phi(x)=\sum\limits_{\ell m}\int\limits_0^\infty \td k_\rC \left(\psi^\rC_{k_\rC\ell m}(x) \hat c_{k_\rC \ell m}+\overline{\psi^\rC_{k_\rC\ell m}}(x)\hat c^\dagger_{k_\rC\ell m}\right)\, ,
\end{align}
where the coefficients $\hat c_{k_\rC\ell m}$ and $\hat c_{k_\rC\ell m}^\dagger$ are operator-valued and act like  annihilation and creation operators of the scalar field on some Fock space. Their associated modes are defined by their asymptotic behaviour on $\mH_-^L\cup\mH_-^+$,
\begin{align}
     \psi^\rC_{k_{\rC}\ell m}\sim \sqrt{\frac{\chi}{4\pi\abs{k_{\rC}}(r_c^2+a^2)}}Y_{\ell m}(\theta, \vp_-) e^{-ik_{\rC}V_-}\, .   
\end{align}
The comparison state is then defined to be the ground state of that Fock space, i.e. the state satisfying $\hat c_{k_{\rC}\ell m}\vert \rC \rangle=0$ for all $k_{\rC}$, $\ell$ and $m$.

We can now combine these results to derive a mode-sum formula for the RSET on $\mH^L_-$, which implies the result on $\mH^R_-$ by the stationarity of the spacetime and states involved. We begin by rewriting the Unruh-modes in terms of Boulware modes, as in the formula for the symmetrized two-point function given in Eq.~\eqref{eq:twopt} of the main text. The behaviour of the Boulware modes near $r=r_-$, combined with the simplified formula for the differential operator $D_{vz'}$ in \eqref{eq:Dsimple}, reveals that the terms proportional to $g_{\nu\vp_-}$ in the Cauchy-horizon limit of the $v\vp_-$-component of the stress-energy tensor lead to a contribution with rapid oscillations in $\omega_-$ of the form $\sim e^{\pm i\omega_- r_*}$. This term will drop out upon integration in the Cauchy-horizon limit.

For the comparison state, we decompose the modes $\phi^\rC_{k_{\rC}\ell m}$, restricted to region $\rII$, into two sets of Boulware-type modes, one set which behaves like $e^{-i\omega_-v}$ on $\mH_-^L$ and vanishes on $\mH_-^R$, and another set which behaves like $e^{-i\omega_- u}$ on $\mH_-^R$ and vanishes on $\mH_-^L$. This decomposition proceeds along the same lines as the computation for the Unruh modes. Once the limit onto $\mH_-^L$ is taken, only the contribution from the first set of modes remains. Moreover, the contribution of the term proportional to $g_{v\vp_-}$ in the $v\vp_-$-component of the expectation value of the stress-energy tensor in the comparison state vanishes also for the comparison state in the Cauchy-horizon limit.

Before we take the limit onto the IH, we rewrite the resulting mode-sum expression in terms of Boulware modes for the difference of expectation values between the Unruh- and comparison state of the $vv$- and $v\vp_-$-components of the quantum stress-energy tensor.

First, we want to avoid an IR divergence at $\omega_-=0$ in the mode-sum formula for the $v\vp_-$-component
\footnote{The mode-sum formula for the $vv$- (or $uu$- or $uv$-) component of the stress-energy tensor  at the IH contains an additional power of $\omega_-$, so that this IR issue is absent for these components.}
which appears when the Cauchy-horizon limit is taken due to a linear divergence in both the scattering coefficients $\mathcal{T}^{\ri/\ru,\rII}_{\omega\ell m}$ and $\mathcal{R}^{\ri/\ru,\rII}_{\omega\ell m}$\footnote{
It can be seen from \eqref{eq: Wronskian int} that at least one of the interior scattering coefficients must diverge at $\omega_-=0$. The fact that both of them diverge linearly can be 
derived from relating the general asymptotic solution of the radial equation near $r_-$ in the case $\omega_-=0$ to an expansion of $R^{\ri/\ru,\rII}_{\omega\ell m}$ near $r_-$ in $\omega_-$. 
\label{fn:IR div}}. 
To avoid this, before taking the Cauchy-horizon limit, we use the behaviour of the Boulware modes under the simultaneous sign flip $(\omega, m)\to (-\omega,-m)$ in order to change the integration range (as well as the integration variable) of the integral over the frequency in the mode-sum formula from $\omega$ ranging over all of $\bR$ to $\omega_-$ ranging over $\bR_+$. Moreover, using the same symmetry, we manipulate the sum over $m$ to run only over positive $m$. This brings the integral over the frequency $\omega_-$ into a form in which, upon taking the Cauchy-horizon limit, obtains a manifestly finite integrand, even at $\omega_-=0$, for the mode-sum formula of the $v\vp_-$-component. The integral over this finite integrand corresponds to the principle value of the Cauchy-horizon limit taken without this rewriting. In order to keep notation consistent, we rewrite the formula for the $vv$-component in the same way. Therefore, in the numerical computations presented in the main document, as well as in the following detailed description of these computations, we consider $\omega_-$ and $m$ as the independent variables, and thus view $\omega=\omega_-+m\Omega_-$, $\omega_+=\omega_-+m(\Omega_--\Omega_+)$ and $\omega_c=\omega_-+m(\Omega_--\Omega_c)$ as functions of $\omega_-$ and $m$. We suppress the functional dependence for brevity and readability.

In the next step, we take the limit of the mode-sum expression for the stress-energy tensor onto the horizon $\mH_-^L$ by taking the limit $u\to -\infty$ while keeping $v$ fixed. This leaves us with a mode-sum expression for the $vv$- or $v\vp_-$-component of the stress-energy tensor on $\mH_-^L$, and therefore $\mH_-^R$ by the $t$-invariance of the states involved.  
Lastly, we write all scattering coefficients in terms of the ones for the ``up,I"- modes and ``in,II"-modes using the Wronskian relations. The final result for the desired components of the RSET in the Unruh state at the IH is given in Eq.~\eqref{eq: formula SET} and Eq.~\eqref{eq:SET integrand} of the main document. 
\subsection{Numerical results}
\label{sec: add num}
In this section, we discuss the implementation of the numerical computation of the RSET at the IH, evaluating Eq.~\eqref{eq: formula SET} in the main document.
In order to evaluate this equation numerically, there are two main tasks. The first one is to solve the angular Teukolsky equation \eqref{eq: angular TE} to obtain the spheroidal harmonics and determine the angular eigenvalues $E^\omega_{\ell m}$. The second one is to determine the scattering coefficients $\mathcal{R}^{\ru,\rI}_{\omega\ell m}$, $\mathcal{R}^{\ri,\rII}_{\omega\ell m}$ and $\mathcal{T}^{\ri,\rII}_{\omega\ell m}$ from the radial Teukolsky equation \eqref{eq: radial TE}. 

We solve the first task by rewriting \eqref{eq: angular TE} in the form of a Heun equation \cite{Suzuki:1998}, and relating the condition that the spheroidal wave function should be regular at both $\theta=0$ and $\theta=\pi$ to the zeroes of a particular Wronskian relation \cite{Hatsuda:2020}. 

For fixed $\ell$ and $m$, we solve for $E^\omega_{\ell m}$ numerically. In the regime where $a\omega$ and $a^2\Lambda/3$ are both small, we  utilized the small-$\omega$ and small-$\Lambda$ expansion for $E^\omega_{\ell m}$ given in \cite{Suzuki:1998} as the seed for the numerical search. In order to obtain the eigenvalues from close to $\omega_-=0$ up to a large, maximum value $\omega_{-,\text{max}}$ of $\omega_-$,  
we iterate over increasing or decreasing $\omega$ starting from $\omega=\omega_0$ with a small stepsize 
$\delta\omega\equiv \Omega_-/n$, 
with a suitably chosen $n$ as specified below, using the previous result as the starting value for the numerical search, up to the $m$-dependent maximum  $\omega=\omega_{-,\text{max}}+m\Omega_-$ for all $m$ and, specifically for $m<0$, down to the minimum $\omega=m\Omega_-$. To avoid $\omega_-=0$, see footnote \ref{fn:IR div}, we do not start from $\omega=0$, but from a small offset $\omega_0$. Lastly, we iterate over $\ell$ and $m$, and use the invariance of $E^\omega_{\ell m}$ under simultaneous sign flips of $\omega$ and $m$ to obtain the angular eigenvalues $E^\omega_{\ell m}$ for all $\ell$, $m$, and $\omega_-$ that are taken into account in the final summation and integration.

\begin{figure}
    \centering
    \includegraphics[scale=0.5]{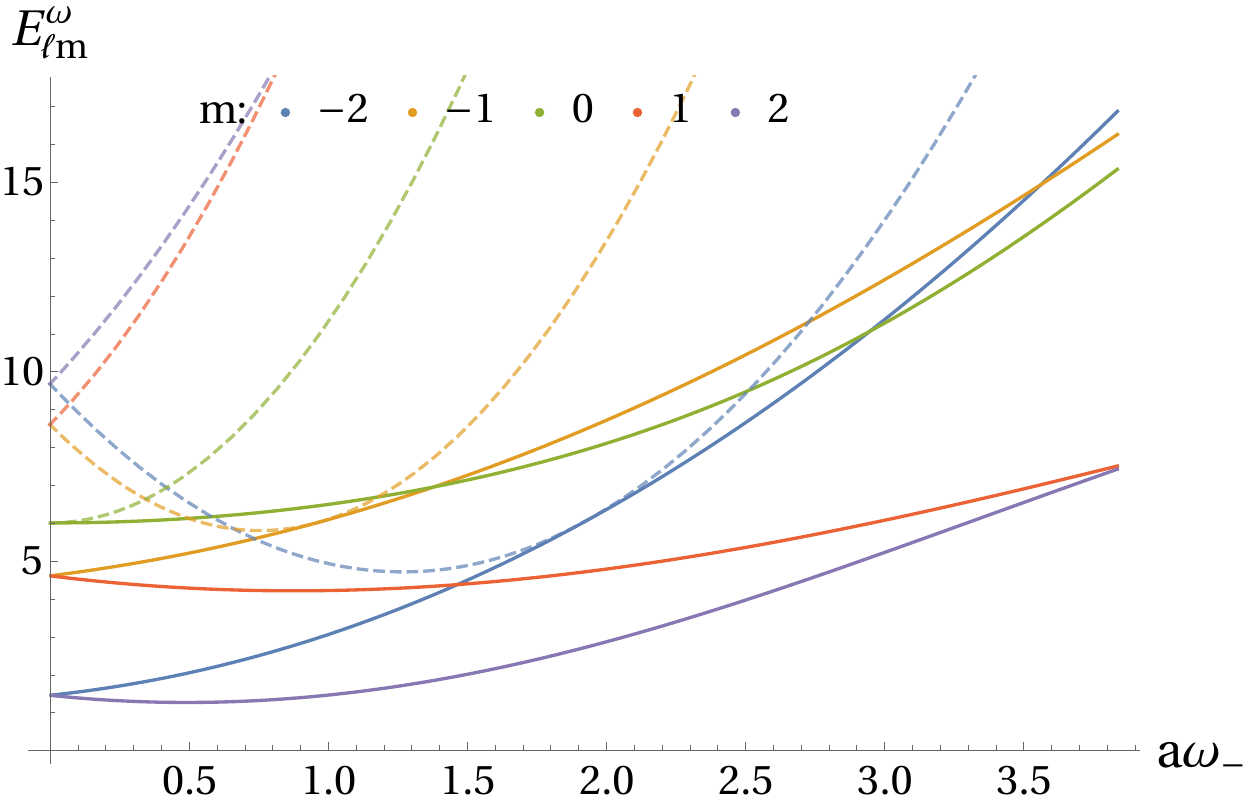}
    \caption{The $a\omega_-$- dependence of the angular eigenvalues $E^\omega_{\ell m}$ for $\ell=2$ and $m=-2,\dots , 2$. The spacetime parameters are  $a=0.4$ and $\Lambda=1/30$. The dashed lines represent the  approximation for the eigenvalue given in \cite[Eq.(4.18)]{Suzuki:1998} valid for $\omega_-\approx -m\Omega_-$
   and small-$\Lambda$.}
    \label{fig: angular EV}
\end{figure}

An example of the $\omega$-dependence of $E^\omega_{\ell m}$ is given in Fig.~\ref{fig: angular EV}. We tested the precision of the angular eigenvalues by checking the linear dependence of the two regular local solutions and find that it is satisfied to at most one digit less than the precision set  in the computation of the angular eigenvalue. We set that precision to vary depending on the parameters but it was always at least 30 digits.
In  Fig.~\ref{fig: angular EV} we compare our numerics for the eigenvalue to
the  small-$\omega$ and small-$\lambda$ expansion in \cite{Suzuki:1998} mentioned above that we used as a seed.
Note that small-$\omega$ corresponds to $\omega_-\approx -m\Omega_-$ in the figure.

Once we have calculated the angular eigenvalues, it is  relatively straightforward to obtain the angular eigenfunctions by using the ansatz described in \cite{Suzuki:1998} or \cite{Hatsuda:2020} and combining it with the regular local solutions to the Heun equation that enter the Wronskian relation. The only remaining difficulty is the normalization, which is obtained by numerical integration over $\theta:0\to \pi$ of the so-obtained angular eigenfunctions. This allows us to make sure that the normalization in Eq.~\eqref{eq:norm S} is satisfied.

Given the angular eigenvalues, one can now solve for the scattering coefficients in the same way as described in \cite{Hollands:2020} and used in \cite{Klein:2021, Klein:2021a}. The radial Teukolsky equation \eqref{eq: radial TE} can be rewritten as a Heun equation as well \cite{Suzuki:1998, Suzuki:1999}, at least in the case $\mu^2=2\Lambda/3$. In this case, one can therefore find local solutions around each of the horizons which can serve as building blocks for the Boulware modes. Matching the two definitions of the Boulware modes arising from the two asymptotic limits in the overlap of their domains then determines the scattering coefficients. 

\begin{figure*}
\begin{subfigure}{0.45\textwidth}
\includegraphics[scale=0.35]{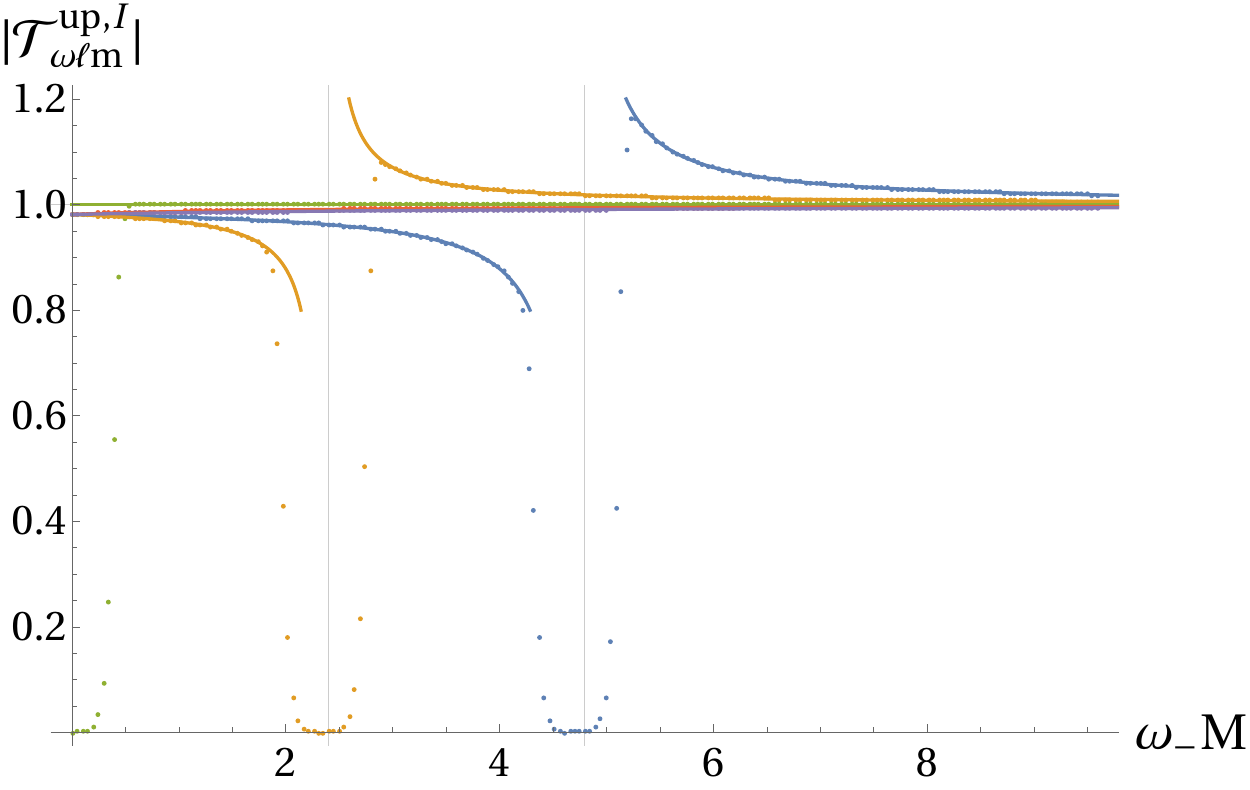}
\end{subfigure}
   \begin{subfigure}{0.45\textwidth}
\includegraphics[scale=0.35]{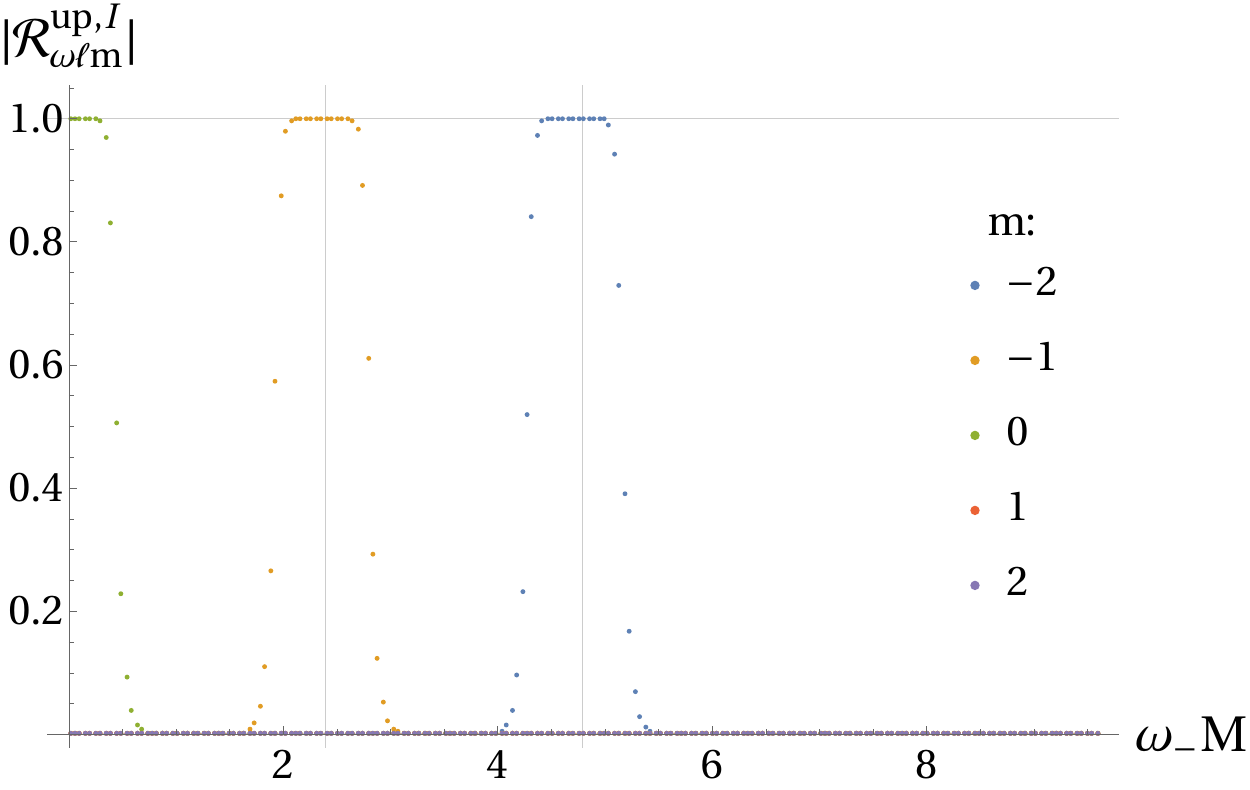}
\end{subfigure}
\begin{subfigure}{0.45\textwidth}
\includegraphics[scale=0.35]{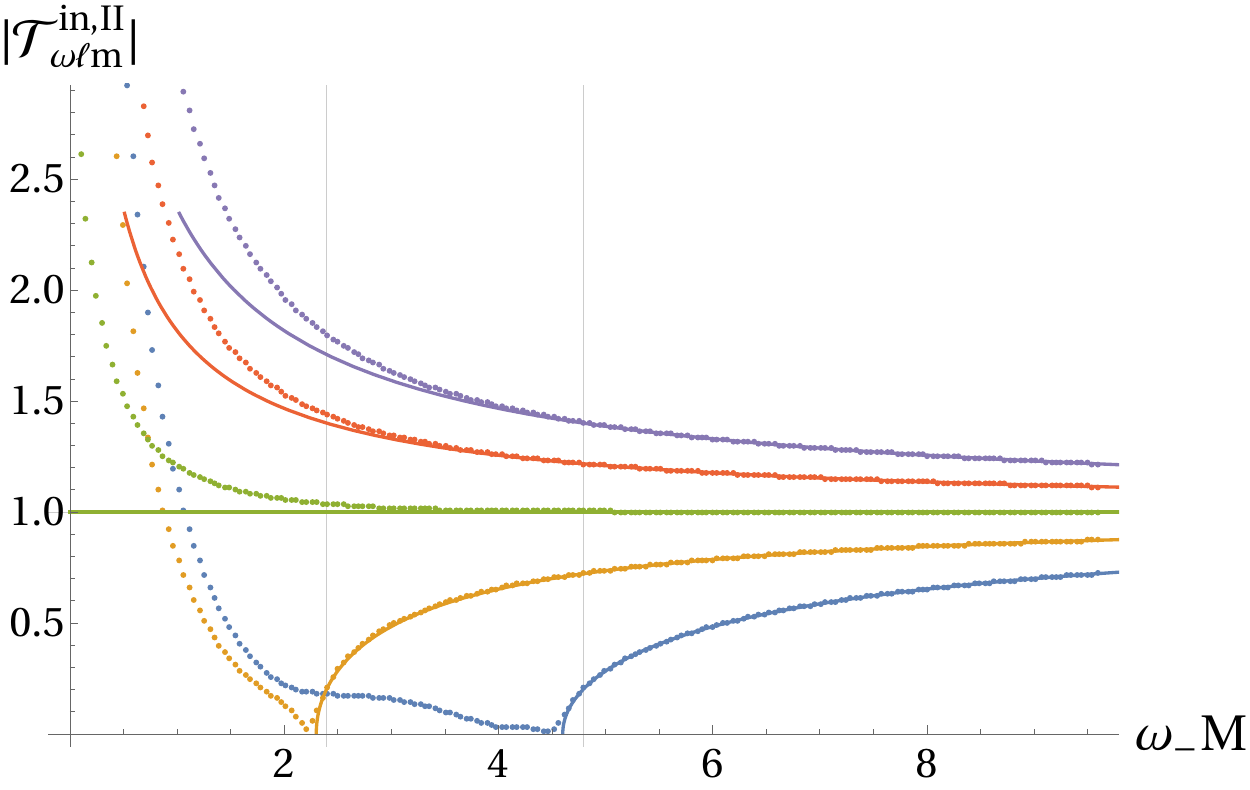}
\end{subfigure}
\begin{subfigure}{0.45\textwidth}
\includegraphics[scale=0.35]{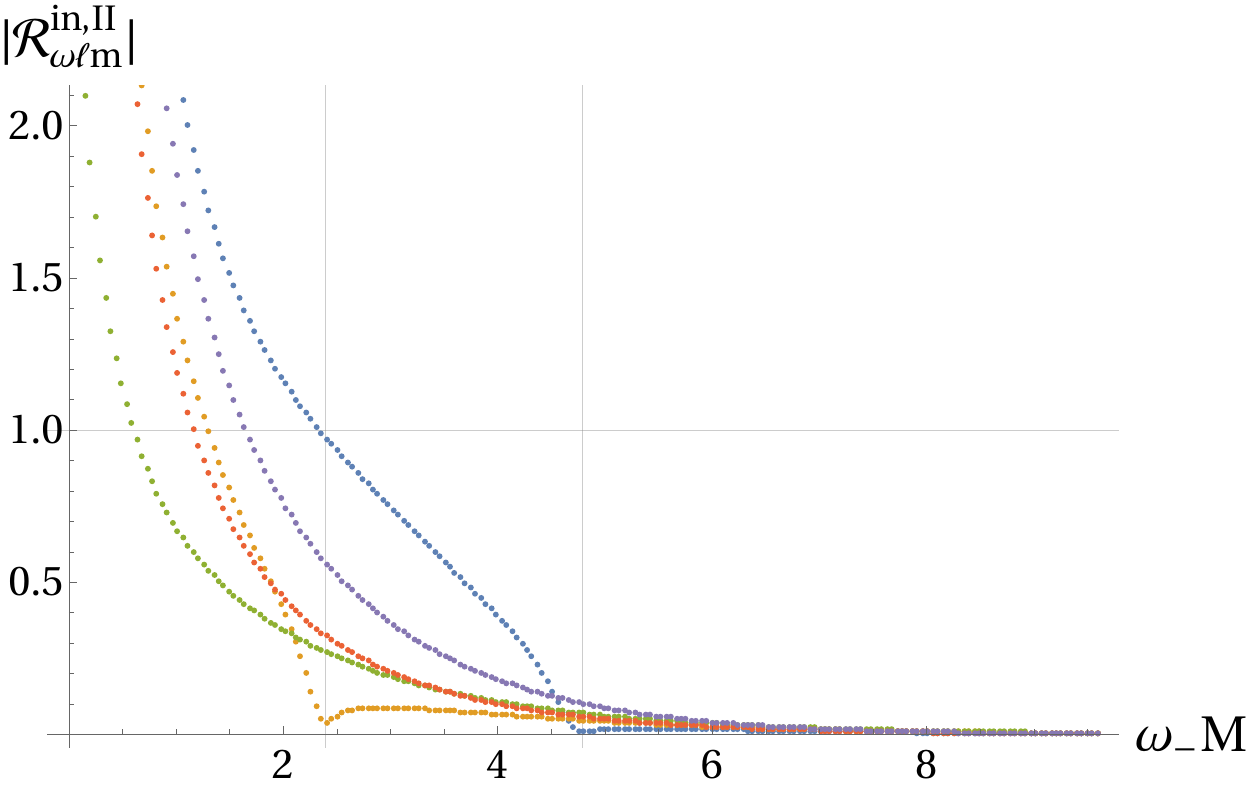}
\end{subfigure}
\caption{The $\omega_-$ dependence of the scattering coefficients for $\ell=2$ for all values of $m$ as indicated in the upper right plot. The spacetime parameters are $a=0.4$ and $\Lambda=1/30$. The vertical gray lines are at $\Omega_-$ and $2\Omega_-$. The transmission coefficients are in excellent agreement with the large-$\omega$ asymptotics shown by the solid lines of the same colour.}
\label{fig:Scat Coeff}
\end{figure*}

As a test of the precision of the scattering coefficient, one can check the Wronskian relations \eqref{eq: Wronskian ext} and \eqref{eq: Wronskian int}. We find that our scattering coefficients satisfy the relations up to 50 digits. Note that this high precision in the scattering coefficients is necessary, since cancellations in the computation of the integrand defined in Eq.~\eqref{eq:SET integrand} of the main paper will lead to a loss of precision, especially the computation of $F^{\vp_-}_{\ell m}(-\omega_-)$ for large values of $\ell$. One can also check that the scattering coefficients exhibit the expected large-$\omega$ asymptotics, namely $\mathcal{R}^{\ri/\ru,\rI/\rII}_{\omega\ell m}\to 0$, $\abs{\mathcal{T}^{\ru,\rI}_{\omega\ell m}}^2\to \omega_+/\omega_c$, and $\abs{\mathcal{T}^{\ri/\ru,\rII}_{\omega\ell m}}^2\to \omega_+/\omega_-$. This can  be conveniently observed in Fig.~\ref{fig:Scat Coeff}, where the absolute values of the scattering coefficients, as well as the expected large-$\omega$ asymptotics for the transmission coefficients, are plotted. 

\begin{figure*}
\begin{subfigure}{0.45\textwidth}
\includegraphics[scale=0.35]{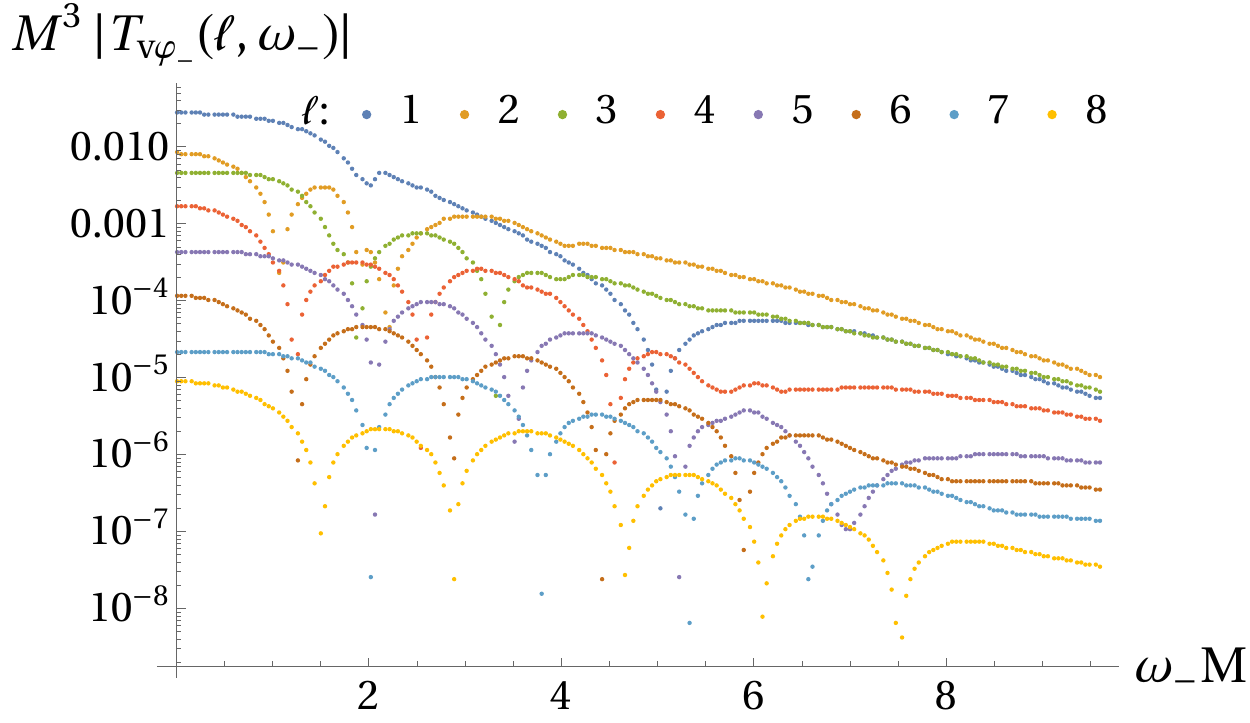}
\label{fig: conv 1}
\end{subfigure}
   \begin{subfigure}{0.45\textwidth}
\includegraphics[scale=0.35]{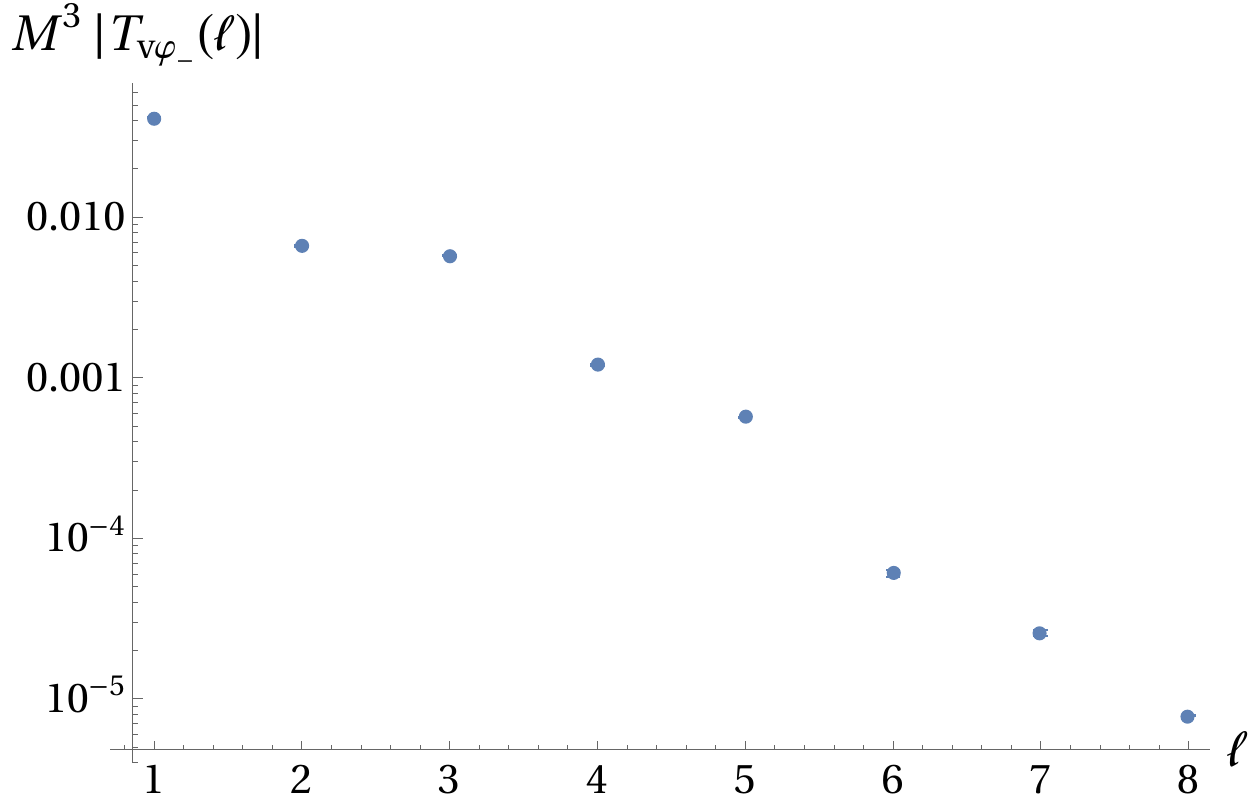}
\label{fig: conv 2}
\end{subfigure}

\caption{
Modes of the angular momentum flux expectation value $\VEV{T_{v\vp_-}}_\rU^{\mathrm{IH}}$ at the IH:
$T_{v\varphi_-}(\ell,\omega_-)$
in terms of $\omega_-$ for fixed $\ell$ (left), and $T_{v\vp_-}(\ell)$ in terms of $\ell$ after integration over $\omega_-$ (right). The spacetime parameters are $a=0.4$ and $\Lambda=1/30$, and the latitude is $\cos\theta=0.9$. We have taken the absolute values to present the results on a logarithmic scale. The modes with $\ell=0$ do not contribute to $\VEV{T_{v\vp_-}}_\rU^{\mathrm{IH}}$ and are therefore not displayed.
Exponential convergence is exhibited both in the left plot
as a function of $\omega_-$ and in the right plot 
as a function of $\ell$.
}
\label{fig: conv}
\end{figure*}

It then only remains to plug all these results into the integrand defined in Eq.~\eqref{eq:SET integrand} of the main document, interpolate in $\omega_-$, and compute the sums and integral. Since in practical calculations one must cut off both the $\ell$-sum and the $\omega_-$-integral at finite values ($\ell_{\text{max}}$  and $\omega_{-,\text{max}}$, respectively), it is very important to understand the convergence of the summation and integration and to choose the cutoffs accordingly, see also the discussion in \cite{Hollands:2020} or \cite{Klein:2021}. To illustrate this, we plot examples of
\begin{align}
    T_{v\varphi_-}(\ell,\omega_-)\equiv \sum\limits_{m=1}^\ell\frac{\chi\left(F^{\vp_-}_{\ell m}(\omega_-)-F^{\vp_-}_{\ell m}(-\omega_-)\right)}{4\pi (r_-^2+a^2)\omega_-}
\end{align}
for fixed $\ell$ as a function of $\omega_-$, as well as
\begin{align}
    T_{v\vp_-}(\ell)\equiv \int\limits_0^{\omega_{-,\text{max}}} \td\omega_-\, T_{v\vp_-}(\ell,\omega_-)
\end{align}
as a function of $\ell$ in Fig.~\ref{fig: conv}.

In the present case, the convergence in $\omega_-$ for fixed $\ell$ and $m$ is exponential, as illustrated in the left panel of Fig.~\ref{fig: conv}. We take the integral over $\omega_-$ for fixed $\ell$ by interpolation and numerical integration. Moreover, the contributions of the individual $\ell$-modes also decay exponentially, see the right panel of  Fig.~\ref{fig: conv}. However, the decay depends on $\theta$: close to the axis of rotation, the regime of exponential decay in $\ell$  starts almost immediately from $\ell=1$ as displayed in Fig.~\ref{fig: conv}, while it starts at a  value of $\ell$ that becomes higher the closer  $\theta$ is to the equatorial plane. This dependence is weaker in the near-extremal regime of large $a$, allowing us to compute $\VEV{\hat T_{v\vp_-}}_\rU^{\rm IH}$ for any value of $\theta$ in this limit. The results are shown in Fig.~\ref{fig:Tvp tdep} of the main document.

Estimating the error due to the cutoff at $\omega_{-,\text{max}}=4\Omega_-$ ($\ell<9$, $a\neq 0.975$) or $11\Omega_-$ (large $\ell$ or $a=0.975$), the offset of $\omega_{0}=1/100$ ($\ell<9$, $a\neq 0.975$) or $1/1000$ (large $\ell$ or $a=0.975$), and the cutoff in $\ell$ at $\ell_{\text{max}}=8$ ($a=a_{\text{max}}-1/200$), $11$ ($a=0.975$, $a=1$), or $13$ ($a=0.95$), we find that near the equatorial plane $\cos\theta=0$, the dominant contribution to the error comes from the cutoff in $\ell$, which is computationally very expensive to increase. In contrast, near the rotation axis, the dominant contribution to the error in this near-extremal regime comes from the offset $\omega_0$, especially in the $\ell=1$ mode.

\begin{figure}
    \centering
    \includegraphics[scale=0.5]{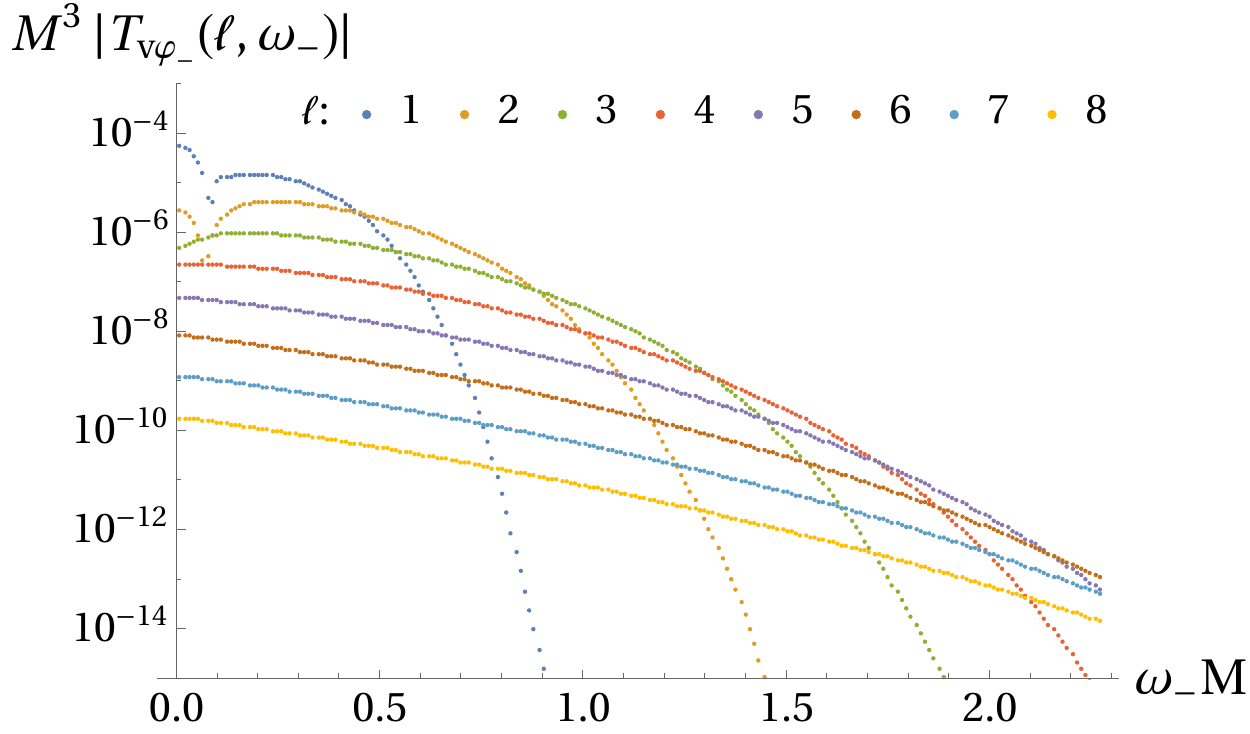}
    \caption{ Absolute values of the modes of the angular momentum flux expectation value $\VEV{T_{v\vp_-}}_\rU^{\mathrm{IH}}$ at the IH in terms of $\omega_-$ for fixed $\ell$ for $\cos \theta=0.9$, $a=1$, and $\Lambda=1/30$. The integrand, especially for $\ell=1$, is concentrated around $\omega_-=0$, and begins rapidly decreasing at comparatively small values of $\omega_-$.}
    \label{fig: conv1 ne}
\end{figure}

We observe that 
for such large $a$ close to extremality, the integrand, especially for $\ell=1$, is  concentrated near $\omega_-=0$, and starts rapidly decaying at comparably small values of $\omega_-$, see Fig.~\ref{fig: conv1 ne} in comparison to the left panel of Fig.~\ref{fig: conv}. As a result, the dominant contribution to the error stems from the offset $\omega_0$ away from $\omega_-=0$. 

To perform the angular integral described in Eq.~\eqref{eq:integral over theta and phi} of the main document, we use the normalization of the spheroidal wave functions, \eqref{eq:norm S}, and interchange the integral over $\theta$ with the summation over $\ell$ and the $\omega_-$-integral. As a result, the absolute value squared of the spheroidal wave functions in the integrand defined in Eq.~\eqref{eq:SET integrand} of the main text simply drops out, as well as the overall factor of $\chi/(2\pi(r_-^2+a^2))$ in Eq.~\eqref{eq: formula SET} of the main text, which cancels with the prefactor of the integral arising from the determinant of the induced metric. The results are again presented in the main text.

\begin{figure}
    \centering
    \includegraphics[scale=0.5]{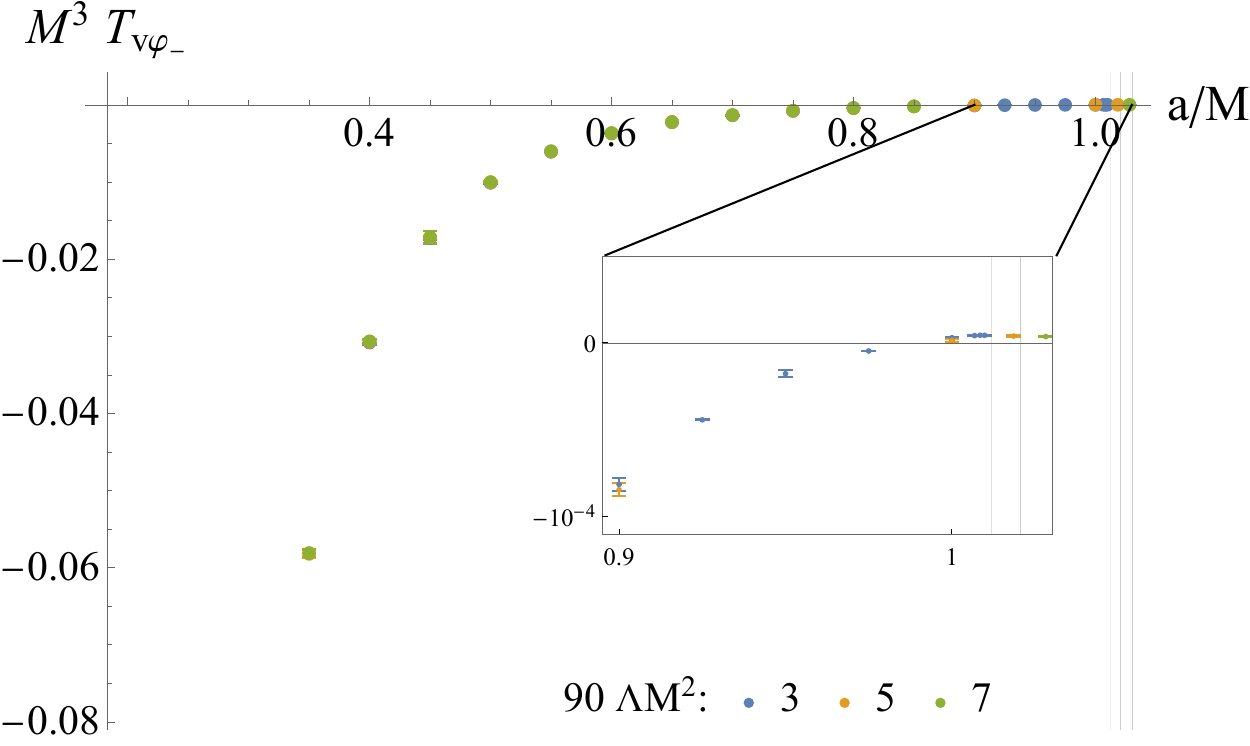}
    \caption{$\VEV{\hat T_{v\vp_-}}_\rU^{\rm IH}$ at $\cos\theta=0.9$  for different  values of $\Lambda$ as a function of  $a$. The three vertical lines represent, from left to right, the extremal values of $a$ for $\Lambda =1/30$, $1/18$, and $ 7/90$. The inlay zooms in  near extremality.}
    \label{fig: Tvp}
\end{figure}

Due to the increase in the number of $\ell$  modes required for small values of $\cos\theta$, especially for smaller values of $a$, we compute the $v\vp_-$- component of the RSET for a larger range of $a$ only in a neighbourhood of the rotation axis. 
In Fig.~\ref{fig: Tvp}  we therefore plot the angular momentum current  at $\cos\theta=0.9$ as a function of $a$ for different values of  $\Lambda$.

While these results do not give sufficient information to estimate the sign of the integral over $\theta$ as described in Eq.~\eqref{eq:integral over theta and phi} of the main text, these results at least indicate the possibility that the integral may change its sign for small enough values of $a$. Moreover, it signals that the dependence on the cosmological constant is not too strong, at least this close to the axis of rotation. 
We thus believe that the `infinite' twisting phenomenon that we present in this Letter (see, e.g., next section) very probably also takes place for $\Lambda=0$ (this is also supported by noting that the $vv$-component of the RSET for $\Lambda=1/270$ plotted in Fig.~\ref{fig: Tvv axis} of our main text  is {\it qualitatively} similar to the same component, but for $\Lambda=0$, plotted in Figs.~2 and 3 of  \cite{Zilberman:2022a}).

Note that, on the rotation axis, only the $m=0$ modes in the mode sum contribute, so that the $v\vp_-$-component of the RSET vanishes there. However, the $vv$-component may be non-zero on the axis, and the fact that only the $m=0$ modes can contribute significantly reduces the increase of computation time with the increase of the $\ell$-cutoff. This allows us to compute the $vv$-component of the RSET  on the rotation axis very efficiently. The results are presented in Fig.~\ref{fig: Tvv axis} of the main document.

For this plot, we use the settings $n=25$, $\omega_0=1/1000$, 
$\omega_{-,\text{max}}=5\Omega_-$, and $\ell_{\max}=8$. With these settings, we obtain results with two significant digits, except for the last data point in the near-extremal limit and the data points with comparatively small $a$, where we only have one significant digit. The dominant contribution to the error in this case comes from the uncertainty arising due to the stepsize in $\omega_-$, which is two times larger than in the rest of the computations.

\subsection{Infinite twisting}

To appreciate more clearly what our findings mean geometrically, consider a null surface ruled by affinely parameterized null geodesics $\xi(\lambda)$ approaching the IH as $\lambda \to 0$, in the spacetime with backreaction effects taken into account. Let ${\mathcal S}(\lambda)$ be cuts of constant $\lambda$ parameterized 
by polar coordinates $x^A = (\theta, \vp)$, and let $n^\mu$ be a past-directed null vector perpendicular to each ${\mathcal S}(\lambda)$ normalized such that $\dot \xi^\mu n_\mu = 1$. We define $\beta \equiv  (\partial_\nu \varphi) \dot\xi^\mu \nabla_\mu n^\nu$. Then $\beta$ measures the amount of twisting in the $\vp$-direction of the null vectors $n^\mu$ -- i.e., of light rays incident perpendicular to ${\mathcal S}(\lambda)$ -- as we move this cut towards the IH. The restriction of the (backreacted) metric 
to ${\mathcal S}(\lambda)$ is written as 
\begin{equation}
g_{AB} \td x^A \td x^B = \gamma^2[\alpha^{-2} \td \theta^2 + \alpha^{2}(\td \vp + \tau \td \theta)^2]. 
\end{equation}
Then $\gamma^2=\sqrt{{\rm det} g_{AB}}$ is the area element, and $\partial_\lambda\tau$ is a measure of how the intrinsic relative twisting of the latitudes ($\theta=$ constant) changes as we move the cut ${\mathcal S}(\lambda)$ to the IH, i.e., taking $\lambda \to 0$. 

We take our results to indicate that, in a spacetime constructed with the semi-classical Einstein equations, the RSET would behave as 
$T_{\lambda\vp} \equiv \VEV{\hat T_{\mu\nu} \dot \xi^\mu \psi^\nu} \sim F_1(\theta) \lambda^{-1}$, $T_{\lambda\lambda} \equiv \VEV{\hat T_{\mu\nu} \dot \xi^\mu \dot \xi^\nu} \sim F_2(\theta) \lambda^{-2}$ as we approach the IH ($\lambda \to 0$), where $\psi^\mu$ is the axial Killing vector field.  The actual values of $F_1$ and $F_2$ would presumably incorporate the gradual evaporation of the black hole -- modeled, e.g., by a Kerr-Vaydia-like metric outside the black hole -- and would presumably not be the same as those of a fixed KdS black hole. We also assume, for simplicity, that the backreacted spacetime is axi-symmetric. These assumptions are motivated by our studies of the RSET in the Unruh state on the \emph{fixed} KdS background, and on this fixed background, would hold asymptotically as $\lambda \to 0$ for an {\it arbitrary} initial Hadamard state by results of \cite{KleinHintz}  for both $T_{\lambda\lambda}$ and $T_{\lambda\varphi}$. Here, we make these assumptions even on a hypothetical solution to the full semi-classical equations including \emph{backreaction}. Of course, when the RSET becomes Planckian, we no longer trust the semi-classical approximation, so `diverging' should be understood as `becoming Planckian' throughout.

The semi-classical Einstein and Raychaudhuri equations are now analyzed under these assumptions. This analysis shows that
${\mathcal S}(\lambda)$ splits into parts with infinite expansion and contraction of the area element as we approach the IH. Furthermore, we argue that, at the latitude separating infinite overall expansion of ${\mathcal S}(\lambda)$ from infinite overall contraction, we must have either diverging twists $\tau$ or $\beta$, or both. See 
Fig.~\ref{fig:twisting}  in the main text
for a schematic illustration of this phenomenon. Finally, in all cases, we have at least one diverging Weyl tensor component $C^B{}_{\lambda A \lambda} \sim \lambda^{-2}$.

To see this, consider
the Raychaudhuri equation for the shear, and the $\lambda\lambda$- and $\lambda \varphi$-components of the semi-classical Einstein equations which can be rearranged as, respectively,
\begin{subequations}
\begin{equation}
\label{eq:a}
    \partial_\lambda (\gamma^2 \sigma_A{}^B) = -C^B{}_{\lambda A \lambda} 
\gamma^2,
\end{equation}
\begin{equation}
\label{eq:b}
\partial_\lambda^2 \gamma = -( \tfrac{1}{2}\sigma_A{}^B \sigma_B{}^A+ 4\pi \, T_{\lambda\lambda})\gamma, 
\end{equation}
\begin{equation}
\label{eq:c}
    \partial_\theta(\gamma^2 \alpha^{4} \partial_\lambda \tau) - \partial_\lambda (\gamma^2 \beta) = \, 8\pi \, \gamma^2 T_{\lambda\vp},
\end{equation}
\begin{equation}
\label{eq:d}
\sigma_A{}^B \sigma_B{}^A = \, 2[4(\partial_\lambda \log \alpha)^2+ \alpha^4(\partial_\lambda \tau)^2].
\end{equation}
\end{subequations}
Here, $\sigma_A{}^B$ is the shear of our family of topological spheres. To estimate the behavior as we approach the IH, i.e., as $\lambda \to 0$, we assume furthermore that any quantity under consideration asymptotically behaves as $f(\theta)\lambda^q$ (or a sum of such terms) for some powers $q$ (or $\log \lambda$), 
as will turn out to be consistent with these equations.

For a latitude $\theta$ such that $T_{\lambda\lambda}>0$, Eq. \eqref{eq:b}
immediately tells us that $\gamma \sim \lambda^q, 0<q<1$, i.e. we have an infinite overall contraction of ${\mathcal S}(\lambda)$. However, we generically expect to also have latitudes such that $T_{\lambda\lambda}<0$, which we have checked numerically e.g. for $a=0.975$ and $\Lambda=1/30$; see \cite{Zilberman:2022a} for similar findings when $\Lambda=0$. If the shear squared term $\sigma_A{}^B \sigma_B{}^A$ does not overwhelm 
$T_{\lambda\lambda}<0$ in Eq. \eqref{eq:b} and if the  
expansion/contraction of the congruence is sufficiently 
small in absolute value for some initial ${\mathcal S}(\lambda_0)$
-- as would be the case, e.g. if this ${\mathcal S}(\lambda_0)$ were taken sufficiently close 
enough to the outer (apparent) horizon -- then Eq. \eqref{eq:b} implies that $\gamma$ is a linear combination of solutions behaving approximately as $\sim \lambda^{q}$ for $q<0$ respectively $q>1$. In the absence of fine tuning there will be a component of the first solution, which then dominates
the expansion as $\lambda \to 0$. Thus, we have an infinite overall expansion. 
We may assume the shear squared term to be small because it should be ``higher order'' in the absence of classical matter source, in the sense that its leading pieces should be incorporated in the quantum RSET for the gravitational perturbations in the first place. Separating the latitudes $\theta = {\rm const.}$ of infinite expansion and infinite contraction, 
we have latitude(s) with zero expansion, i.e., $\gamma \sim \lambda^0$.
Furthermore, unless $T_{\lambda\lambda}$ is zero precisely at such a latitude -- which presumably would require an infinite fine-tuning -- the two terms on the r.h.s. of Eq. \eqref{eq:b} are both non-zero and scale in exactly the same way, i.e. $-T_{\lambda\lambda} \sim \sigma_A{}^B \sigma_B{}^A \lesssim \lambda^{-2}$.

Now we look in detail at such a latitude with zero expansion separating regions of infinite expansion and infinite contraction of our spheres ${\mathcal S}(\lambda)$ approaching the IH ($\lambda \to 0$). From Eq. \eqref{eq:a}
we find that at least one component of $C^B{}_{\lambda A \lambda} \sim \lambda^{-2}$, i.e. we have a curvature singularity at the IH. 
From Eq. \eqref{eq:d}, we get 
i. $\alpha \sim \lambda^p$ for $p>0$ or $p<0$ and/or ii. $\alpha^2 \partial_\lambda \tau \sim \lambda^{-1}$. From Eq. \eqref{eq:c} we get 
iii. $\beta \sim \log \lambda$ and/or iv. $\alpha^4 \partial_\lambda \tau \sim \lambda^{-1}$ (by integrating eq. \eqref{eq:c} from $0$ to $\theta$). 

Suppose we are in case i. with $p>0$. Then if iv. holds, 
we find that $\partial_\lambda \tau$ is diverging, and if iii. holds, 
$\beta$ is diverging. In case i. with $p<0$, Eq. \eqref{eq:d} and 
$\sigma_A{}^B \sigma_B{}^A \sim \lambda^{-2}$ is telling us that 
$\alpha^2 \partial_\lambda \tau \lesssim \lambda^{-1}$, so 
$\partial_\lambda \tau \lesssim \lambda^{-1-2p}$. Now we consider Eq. \eqref{eq:c} as an equation for $\beta$. Since 
$T_{\lambda \varphi} \sim \lambda^{-1}$, we will have $\beta \sim \log \lambda$ unless there is an exact cancellation of $T_{\lambda \varphi} \sim \lambda^{-1}$ with the first term on the l.h.s. of Eq. \eqref{eq:c}.
This would mean that $\partial_\lambda \tau \sim \lambda^{-1-4p}$.
We shall now make an estimate of $|p|$ under this assumption, which we note should be a highly fine-tuned case anyhow. 

First we note that since $\alpha \sim \lambda^p$ and $p<0$, the second term on the r.h.s. in Eq. \eqref{eq:d}
is subdominant to the first term on the r.h.s., so 
$\sigma_A{}^B \sigma_B{}^A \sim 8p^2 \lambda^{-2}$ with this precise pre-factor. In the present case of a latitude with zero expansion, the r.h.s. of Eq. \eqref{eq:b} is zero to leading order, 
so this term must be cancelled precisely by $T_{\lambda\lambda}$ for $\lambda \to 0$. This is saying that 
\begin{equation}
    p= \lim_{\lambda \to 0} \left[-\lambda \sqrt{\tfrac{\pi}{2} |T_{\lambda\lambda}|} \right] = -\tfrac{1}{\kappa_-}
\sqrt{\tfrac{\pi}{2}|T_{vv}^{\rm IH}|} 
\sim -\frac{1}{M} \frac{a_{\rm max}}{(a_{\rm max}-a)}
\sqrt{|M^4 T_{vv}^{\rm IH}|},
\end{equation}
where we assumed in the last step that we are close to extremality, as is the case for the simulations carried out in this work. Since we are working in geometrical 
units, $M$ is a length scale in units of the Planck length. For a Kerr-Vaydia-like spacetime describing a slowly evaporating black hole, we think of this scale $M$ as locally describing the size of the evaporating black hole, at least sufficiently close to the outer (apparent) horizon. Consequently, at small $\lambda \to 0$, this $M$
could effectively be much smaller than that of the original black hole though still 
comfortably larger than $O(1)$ in order for the semi-classical approximation to apply. 

We now substitute our numerical results for $|M^4 T_{vv}^{\rm IH}|$ to obtain an order of magnitude estimate for $p$. In the case considered $M^4T_{vv}^{\rm IH}<0$, so to get an upper bound on $|p|$ we take the most negative value for this quantity. E.g., for $a/M=0.975$ and $M^2\Lambda=1/30$ (as in the orange plot of Fig.~\ref{fig:Tvp tdep} of the main text), a numerical calculation of precisely the same nature as described in this Supplement yields $M^4T_{vv}^{\rm IH} \sim -6.5 \cdot 10^{-7}$. Since we should take $O(1) \lesssim M$ close to the IH, this leads to $|p| \lesssim 0.01$. Since $\partial_\lambda \tau \sim \lambda^{-1-4p}$, this would be diverging with this estimate for $|p|$.

Suppose finally we are in case ii. but not i. Then $\alpha = \lambda^0$ and, again, $\partial_\lambda  \tau$ is diverging. In summary, we find evidence that, at a latitude separating infinitely expanding and contracting cross section surface area of a cut approaching the IH along a congruence of lightlike geodesics, either $\partial_\lambda \tau$ or $\beta$ or both are diverging. Both quantities are a measure of a relative twisting, see Fig.~\ref{fig:twisting} in the main text.




\begin{thebibliography}{47}%
\makeatletter
\providecommand \@ifxundefined [1]{%
 \@ifx{#1\undefined}
}%
\providecommand \@ifnum [1]{%
 \ifnum #1\expandafter \@firstoftwo
 \else \expandafter \@secondoftwo
 \fi
}%
\providecommand \@ifx [1]{%
 \ifx #1\expandafter \@firstoftwo
 \else \expandafter \@secondoftwo
 \fi
}%
\providecommand \natexlab [1]{#1}%
\providecommand \enquote  [1]{``#1''}%
\providecommand \bibnamefont  [1]{#1}%
\providecommand \bibfnamefont [1]{#1}%
\providecommand \citenamefont [1]{#1}%
\providecommand \href@noop [0]{\@secondoftwo}%
\providecommand \href [0]{\begingroup \@sanitize@url \@href}%
\providecommand \@href[1]{\@@startlink{#1}\@@href}%
\providecommand \@@href[1]{\endgroup#1\@@endlink}%
\providecommand \@sanitize@url [0]{\catcode `\\12\catcode `\$12\catcode
  `\&12\catcode `\#12\catcode `\^12\catcode `\_12\catcode `\%12\relax}%
\providecommand \@@startlink[1]{}%
\providecommand \@@endlink[0]{}%
\providecommand \url  [0]{\begingroup\@sanitize@url \@url }%
\providecommand \@url [1]{\endgroup\@href {#1}{\urlprefix }}%
\providecommand \urlprefix  [0]{URL }%
\providecommand \Eprint [0]{\href }%
\providecommand \doibase [0]{http://dx.doi.org/}%
\providecommand \selectlanguage [0]{\@gobble}%
\providecommand \bibinfo  [0]{\@secondoftwo}%
\providecommand \bibfield  [0]{\@secondoftwo}%
\providecommand \translation [1]{[#1]}%
\providecommand \BibitemOpen [0]{}%
\providecommand \bibitemStop [0]{}%
\providecommand \bibitemNoStop [0]{.\EOS\space}%
\providecommand \EOS [0]{\spacefactor3000\relax}%
\providecommand \BibitemShut  [1]{\csname bibitem#1\endcsname}%
\let\auto@bib@innerbib\@empty
\bibitem [{\citenamefont {Penrose}(1974)}]{Penrose:1974}%
  \BibitemOpen
  \bibfield  {author} {\bibinfo {author} {\bibfnamefont {R.}~\bibnamefont
  {Penrose}},\ }\enquote {\bibinfo {title} {Gravitational radiation and
  gravitational collapse},}\ \ (\bibinfo  {publisher} {Springer},\ \bibinfo
  {address} {Heidelberg},\ \bibinfo {year} {1974})\ Chap.\ \bibinfo {chapter}
  {Gravitational collapse}\BibitemShut {NoStop}%
\bibitem [{\citenamefont {Poisson}\ and\ \citenamefont
  {Israel}(1989)}]{Poisson:1989}%
  \BibitemOpen
  \bibfield  {author} {\bibinfo {author} {\bibfnamefont {E.}~\bibnamefont
  {Poisson}}\ and\ \bibinfo {author} {\bibfnamefont {W.}~\bibnamefont
  {Israel}},\ }\href {\doibase 10.1103/PhysRevLett.63.1663} {\bibfield
  {journal} {\bibinfo  {journal} {Phys. Rev. Lett.}\ }\textbf {\bibinfo
  {volume} {63}},\ \bibinfo {pages} {1663} (\bibinfo {year}
  {1989})}\BibitemShut {NoStop}%
\bibitem [{\citenamefont {Poisson}\ and\ \citenamefont
  {Israel}(1990)}]{Poisson:1990}%
  \BibitemOpen
  \bibfield  {author} {\bibinfo {author} {\bibfnamefont {E.}~\bibnamefont
  {Poisson}}\ and\ \bibinfo {author} {\bibfnamefont {W.}~\bibnamefont
  {Israel}},\ }\href {\doibase 10.1103/PhysRevD.41.1796} {\bibfield  {journal}
  {\bibinfo  {journal} {Phys.\ Rev.\ D}\ }\textbf {\bibinfo {volume} {41}},\
  \bibinfo {pages} {1796} (\bibinfo {year} {1990})}\BibitemShut {NoStop}%
\bibitem [{\citenamefont {Ori}(1991)}]{Ori:1991}%
  \BibitemOpen
  \bibfield  {author} {\bibinfo {author} {\bibfnamefont {A.}~\bibnamefont
  {Ori}},\ }\href {\doibase 10.1103/PhysRevLett.67.789} {\bibfield  {journal}
  {\bibinfo  {journal} {Phys. Rev. Lett.}\ }\textbf {\bibinfo {volume} {67}},\
  \bibinfo {pages} {789} (\bibinfo {year} {1991})}\BibitemShut {NoStop}%
\bibitem [{\citenamefont {Dafermos}\ and\ \citenamefont
  {Luk}(2017)}]{Dafermos:2017}%
  \BibitemOpen
  \bibfield  {author} {\bibinfo {author} {\bibfnamefont {M.}~\bibnamefont
  {Dafermos}}\ and\ \bibinfo {author} {\bibfnamefont {J.}~\bibnamefont {Luk}},\
  }\href@noop {} {\  (\bibinfo {year} {2017})},\ \Eprint
  {http://arxiv.org/abs/1710.01722} {arXiv:1710.01722 [gr-qc]} \BibitemShut
  {NoStop}%
\bibitem [{\citenamefont {Christodoulou}(2009)}]{Christodoulou:2008}%
  \BibitemOpen
  \bibfield  {author} {\bibinfo {author} {\bibfnamefont {D.}~\bibnamefont
  {Christodoulou}},\ }\href@noop {} {\emph {\bibinfo {title} {{The Formation of
  Black Holes in General Relativity}}}}\ (\bibinfo  {publisher} {European
  Mathematical Society Publishing House},\ \bibinfo {address} {Zürich},\
  \bibinfo {year} {2009})\ \Eprint {http://arxiv.org/abs/0805.3880}
  {arXiv:0805.3880 [gr-qc]} \BibitemShut {NoStop}%
\bibitem [{\citenamefont {Mellor}\ and\ \citenamefont
  {Moss}(1990)}]{Mellor:1990}%
  \BibitemOpen
  \bibfield  {author} {\bibinfo {author} {\bibfnamefont {F.}~\bibnamefont
  {Mellor}}\ and\ \bibinfo {author} {\bibfnamefont {I.}~\bibnamefont {Moss}},\
  }\href {\doibase 10.1103/PhysRevD.41.403} {\bibfield  {journal} {\bibinfo
  {journal} {Phys. Rev. D}\ }\textbf {\bibinfo {volume} {41}},\ \bibinfo
  {pages} {403} (\bibinfo {year} {1990})}\BibitemShut {NoStop}%
\bibitem [{\citenamefont {Mellor}\ and\ \citenamefont
  {Moss}(1992)}]{Mellor:1992}%
  \BibitemOpen
  \bibfield  {author} {\bibinfo {author} {\bibfnamefont {F.}~\bibnamefont
  {Mellor}}\ and\ \bibinfo {author} {\bibfnamefont {I.}~\bibnamefont {Moss}},\
  }\href {\doibase 10.1088/0264-9381/9/4/001} {\bibfield  {journal} {\bibinfo
  {journal} {Classical and Quantum Gravity}\ }\textbf {\bibinfo {volume} {9}},\
  \bibinfo {pages} {L43} (\bibinfo {year} {1992})}\BibitemShut {NoStop}%
\bibitem [{\citenamefont {Brady}\ \emph {et~al.}(1998)\citenamefont {Brady},
  \citenamefont {Moss},\ and\ \citenamefont {Myers}}]{Brady:1998}%
  \BibitemOpen
  \bibfield  {author} {\bibinfo {author} {\bibfnamefont {P.~R.}\ \bibnamefont
  {Brady}}, \bibinfo {author} {\bibfnamefont {I.~G.}\ \bibnamefont {Moss}}, \
  and\ \bibinfo {author} {\bibfnamefont {R.~C.}\ \bibnamefont {Myers}},\ }\href
  {\doibase 10.1103/PhysRevLett.80.3432} {\bibfield  {journal} {\bibinfo
  {journal} {Phys.\ Rev.\ Lett.}\ }\textbf {\bibinfo {volume} {80}},\ \bibinfo
  {pages} {3432} (\bibinfo {year} {1998})},\ \Eprint
  {http://arxiv.org/abs/gr-qc/9801032} {arXiv:gr-qc/9801032} \BibitemShut
  {NoStop}%
\bibitem [{\citenamefont {Dafermos}\ and\ \citenamefont
  {Shlapentokh-Rothman}(2017)}]{Dafermos:2015}%
  \BibitemOpen
  \bibfield  {author} {\bibinfo {author} {\bibfnamefont {M.}~\bibnamefont
  {Dafermos}}\ and\ \bibinfo {author} {\bibfnamefont {Y.}~\bibnamefont
  {Shlapentokh-Rothman}},\ }\href {\doibase 10.1007/s00220-016-2771-z}
  {\bibfield  {journal} {\bibinfo  {journal} {Commun. Math. Phys.}\ }\textbf
  {\bibinfo {volume} {350}},\ \bibinfo {pages} {985} (\bibinfo {year}
  {2017})},\ \Eprint {http://arxiv.org/abs/1512.08260} {arXiv:1512.08260
  [gr-qc]} \BibitemShut {NoStop}%
\bibitem [{\citenamefont {Luk}\ and\ \citenamefont {Oh}(2017)}]{Luk:2015}%
  \BibitemOpen
  \bibfield  {author} {\bibinfo {author} {\bibfnamefont {J.}~\bibnamefont
  {Luk}}\ and\ \bibinfo {author} {\bibfnamefont {S.-J.}\ \bibnamefont {Oh}},\
  }\href {\doibase 10.1215/00127094-3715189} {\bibfield  {journal} {\bibinfo
  {journal} {Duke Math. J.}\ }\textbf {\bibinfo {volume} {166}},\ \bibinfo
  {pages} {437} (\bibinfo {year} {2017})},\ \Eprint
  {http://arxiv.org/abs/1501.04598} {arXiv:1501.04598 [gr-qc]} \BibitemShut
  {NoStop}%
\bibitem [{\citenamefont {Cardoso}\ \emph
  {et~al.}(2018{\natexlab{a}})\citenamefont {Cardoso}, \citenamefont {Costa},
  \citenamefont {Destounis}, \citenamefont {Hintz},\ and\ \citenamefont
  {Jansen}}]{Cardoso:2018}%
  \BibitemOpen
  \bibfield  {author} {\bibinfo {author} {\bibfnamefont {V.}~\bibnamefont
  {Cardoso}}, \bibinfo {author} {\bibfnamefont {J.~L.}\ \bibnamefont {Costa}},
  \bibinfo {author} {\bibfnamefont {K.}~\bibnamefont {Destounis}}, \bibinfo
  {author} {\bibfnamefont {P.}~\bibnamefont {Hintz}}, \ and\ \bibinfo {author}
  {\bibfnamefont {A.}~\bibnamefont {Jansen}},\ }\href {\doibase
  10.1103/PhysRevD.98.104007} {\bibfield  {journal} {\bibinfo  {journal} {Phys.
  Rev. D}\ }\textbf {\bibinfo {volume} {98}},\ \bibinfo {pages} {104007}
  (\bibinfo {year} {2018}{\natexlab{a}})},\ \Eprint
  {http://arxiv.org/abs/1808.03631} {arXiv:1808.03631 [gr-qc]} \BibitemShut
  {NoStop}%
\bibitem [{\citenamefont {Dias}\ \emph {et~al.}(2019)\citenamefont {Dias},
  \citenamefont {Reall},\ and\ \citenamefont {Santos}}]{Dias:2018}%
  \BibitemOpen
  \bibfield  {author} {\bibinfo {author} {\bibfnamefont {O.~J.}\ \bibnamefont
  {Dias}}, \bibinfo {author} {\bibfnamefont {H.~S.}\ \bibnamefont {Reall}}, \
  and\ \bibinfo {author} {\bibfnamefont {J.~E.}\ \bibnamefont {Santos}},\
  }\href {\doibase 10.1088/1361-6382/aafcf2} {\bibfield  {journal} {\bibinfo
  {journal} {Class. Quant. Grav.}\ }\textbf {\bibinfo {volume} {36}},\ \bibinfo
  {pages} {045005} (\bibinfo {year} {2019})},\ \Eprint
  {http://arxiv.org/abs/1808.04832} {arXiv:1808.04832 [gr-qc]} \BibitemShut
  {NoStop}%
\bibitem [{\citenamefont {Dias}\ \emph {et~al.}(2018)\citenamefont {Dias},
  \citenamefont {Eperon}, \citenamefont {Reall},\ and\ \citenamefont
  {Santos}}]{Dias:2018a}%
  \BibitemOpen
  \bibfield  {author} {\bibinfo {author} {\bibfnamefont {O.~J.~C.}\
  \bibnamefont {Dias}}, \bibinfo {author} {\bibfnamefont {F.~C.}\ \bibnamefont
  {Eperon}}, \bibinfo {author} {\bibfnamefont {H.~S.}\ \bibnamefont {Reall}}, \
  and\ \bibinfo {author} {\bibfnamefont {J.~E.}\ \bibnamefont {Santos}},\
  }\href {\doibase 10.1103/PhysRevD.97.104060} {\bibfield  {journal} {\bibinfo
  {journal} {Phys. Rev. D}\ }\textbf {\bibinfo {volume} {97}},\ \bibinfo
  {pages} {104060} (\bibinfo {year} {2018})},\ \Eprint
  {http://arxiv.org/abs/1801.09694} {arXiv:1801.09694 [gr-qc]} \BibitemShut
  {NoStop}%
\bibitem [{\citenamefont {Costa}\ \emph {et~al.}(2018)\citenamefont {Costa},
  \citenamefont {Gir{\~a}o}, \citenamefont {Nat{\'a}rio},\ and\ \citenamefont
  {Silva}}]{Costa:2018}%
  \BibitemOpen
  \bibfield  {author} {\bibinfo {author} {\bibfnamefont {J.~L.}\ \bibnamefont
  {Costa}}, \bibinfo {author} {\bibfnamefont {P.~M.}\ \bibnamefont
  {Gir{\~a}o}}, \bibinfo {author} {\bibfnamefont {J.}~\bibnamefont
  {Nat{\'a}rio}}, \ and\ \bibinfo {author} {\bibfnamefont {J.~D.}\ \bibnamefont
  {Silva}},\ }\href@noop {} {\bibfield  {journal} {\bibinfo  {journal}
  {Communications in Mathematical Physics}\ }\textbf {\bibinfo {volume}
  {361}},\ \bibinfo {pages} {289} (\bibinfo {year} {2018})}\BibitemShut
  {NoStop}%
\bibitem [{\citenamefont {Luna}\ \emph {et~al.}(2019)\citenamefont {Luna},
  \citenamefont {Zilh\~ao}, \citenamefont {Cardoso}, \citenamefont {Costa},\
  and\ \citenamefont {Nat\'ario}}]{Luna:2019}%
  \BibitemOpen
  \bibfield  {author} {\bibinfo {author} {\bibfnamefont {R.}~\bibnamefont
  {Luna}}, \bibinfo {author} {\bibfnamefont {M.}~\bibnamefont {Zilh\~ao}},
  \bibinfo {author} {\bibfnamefont {V.}~\bibnamefont {Cardoso}}, \bibinfo
  {author} {\bibfnamefont {J.~a.~L.}\ \bibnamefont {Costa}}, \ and\ \bibinfo
  {author} {\bibfnamefont {J.}~\bibnamefont {Nat\'ario}},\ }\href {\doibase
  10.1103/PhysRevD.99.064014} {\bibfield  {journal} {\bibinfo  {journal} {Phys.
  Rev. D}\ }\textbf {\bibinfo {volume} {99}},\ \bibinfo {pages} {064014}
  (\bibinfo {year} {2019})},\ \bibinfo {note} {[Addendum: Phys.Rev.D 103,
  104043 (2021)]},\ \Eprint {http://arxiv.org/abs/1810.00886} {arXiv:1810.00886
  [gr-qc]} \BibitemShut {NoStop}%
\bibitem [{\citenamefont {Cardoso}\ \emph
  {et~al.}(2018{\natexlab{b}})\citenamefont {Cardoso}, \citenamefont {Costa},
  \citenamefont {Destounis}, \citenamefont {Hintz},\ and\ \citenamefont
  {Jansen}}]{Cardoso:2017}%
  \BibitemOpen
  \bibfield  {author} {\bibinfo {author} {\bibfnamefont {V.}~\bibnamefont
  {Cardoso}}, \bibinfo {author} {\bibfnamefont {J.~a.~L.}\ \bibnamefont
  {Costa}}, \bibinfo {author} {\bibfnamefont {K.}~\bibnamefont {Destounis}},
  \bibinfo {author} {\bibfnamefont {P.}~\bibnamefont {Hintz}}, \ and\ \bibinfo
  {author} {\bibfnamefont {A.}~\bibnamefont {Jansen}},\ }\href {\doibase
  10.1103/PhysRevLett.120.031103} {\bibfield  {journal} {\bibinfo  {journal}
  {Phys. Rev. Lett.}\ }\textbf {\bibinfo {volume} {120}},\ \bibinfo {pages}
  {031103} (\bibinfo {year} {2018}{\natexlab{b}})},\ \Eprint
  {http://arxiv.org/abs/1711.10502} {arXiv:1711.10502 [gr-qc]} \BibitemShut
  {NoStop}%
\bibitem [{\citenamefont {Hollands}\ \emph
  {et~al.}(2020{\natexlab{a}})\citenamefont {Hollands}, \citenamefont {Wald},\
  and\ \citenamefont {Zahn}}]{Hollands:2019}%
  \BibitemOpen
  \bibfield  {author} {\bibinfo {author} {\bibfnamefont {S.}~\bibnamefont
  {Hollands}}, \bibinfo {author} {\bibfnamefont {R.~M.}\ \bibnamefont {Wald}},
  \ and\ \bibinfo {author} {\bibfnamefont {J.}~\bibnamefont {Zahn}},\ }\href
  {\doibase 10.1088/1361-6382/ab8052} {\bibfield  {journal} {\bibinfo
  {journal} {Class. Quant. Grav.}\ }\textbf {\bibinfo {volume} {37}},\ \bibinfo
  {pages} {115009} (\bibinfo {year} {2020}{\natexlab{a}})},\ \Eprint
  {http://arxiv.org/abs/1912.06047} {arXiv:1912.06047 [gr-qc]} \BibitemShut
  {NoStop}%
\bibitem [{\citenamefont {Hollands}\ \emph
  {et~al.}(2020{\natexlab{b}})\citenamefont {Hollands}, \citenamefont {Klein},\
  and\ \citenamefont {Zahn}}]{Hollands:2020}%
  \BibitemOpen
  \bibfield  {author} {\bibinfo {author} {\bibfnamefont {S.}~\bibnamefont
  {Hollands}}, \bibinfo {author} {\bibfnamefont {C.}~\bibnamefont {Klein}}, \
  and\ \bibinfo {author} {\bibfnamefont {J.}~\bibnamefont {Zahn}},\ }\href
  {\doibase 10.1103/PhysRevD.102.085004} {\bibfield  {journal} {\bibinfo
  {journal} {Phys. Rev. D}\ }\textbf {\bibinfo {volume} {102}},\ \bibinfo
  {pages} {085004} (\bibinfo {year} {2020}{\natexlab{b}})},\ \Eprint
  {http://arxiv.org/abs/2006.10991} {arXiv:2006.10991 [gr-qc]} \BibitemShut
  {NoStop}%
\bibitem [{\citenamefont {Lanir}\ \emph {et~al.}(2018)\citenamefont {Lanir},
  \citenamefont {Levi}, \citenamefont {Ori},\ and\ \citenamefont
  {Sela}}]{Lanir:2017}%
  \BibitemOpen
  \bibfield  {author} {\bibinfo {author} {\bibfnamefont {A.}~\bibnamefont
  {Lanir}}, \bibinfo {author} {\bibfnamefont {A.}~\bibnamefont {Levi}},
  \bibinfo {author} {\bibfnamefont {A.}~\bibnamefont {Ori}}, \ and\ \bibinfo
  {author} {\bibfnamefont {O.}~\bibnamefont {Sela}},\ }\href {\doibase
  10.1103/PhysRevD.97.024033} {\bibfield  {journal} {\bibinfo  {journal} {Phys.
  Rev. D}\ }\textbf {\bibinfo {volume} {97}},\ \bibinfo {pages} {024033}
  (\bibinfo {year} {2018})},\ \Eprint {http://arxiv.org/abs/1710.07267}
  {arXiv:1710.07267 [gr-qc]} \BibitemShut {NoStop}%
\bibitem [{\citenamefont {Zilberman}\ \emph {et~al.}(2020)\citenamefont
  {Zilberman}, \citenamefont {Levi},\ and\ \citenamefont
  {Ori}}]{Zilberman:2019}%
  \BibitemOpen
  \bibfield  {author} {\bibinfo {author} {\bibfnamefont {N.}~\bibnamefont
  {Zilberman}}, \bibinfo {author} {\bibfnamefont {A.}~\bibnamefont {Levi}}, \
  and\ \bibinfo {author} {\bibfnamefont {A.}~\bibnamefont {Ori}},\ }\href
  {\doibase 10.1103/PhysRevLett.124.171302} {\bibfield  {journal} {\bibinfo
  {journal} {Phys. Rev. Lett.}\ }\textbf {\bibinfo {volume} {124}},\ \bibinfo
  {pages} {171302} (\bibinfo {year} {2020})},\ \Eprint
  {http://arxiv.org/abs/1906.11303} {arXiv:1906.11303 [gr-qc]} \BibitemShut
  {NoStop}%
\bibitem [{\citenamefont {Herman}\ and\ \citenamefont
  {Hiscock}(1994)}]{Herman:1994}%
  \BibitemOpen
  \bibfield  {author} {\bibinfo {author} {\bibfnamefont {R.}~\bibnamefont
  {Herman}}\ and\ \bibinfo {author} {\bibfnamefont {W.~A.}\ \bibnamefont
  {Hiscock}},\ }\href {\doibase 10.1103/PhysRevD.49.3946} {\bibfield  {journal}
  {\bibinfo  {journal} {Phys. Rev. D}\ }\textbf {\bibinfo {volume} {49}},\
  \bibinfo {pages} {3946} (\bibinfo {year} {1994})}\BibitemShut {NoStop}%
\bibitem [{\citenamefont {Sorkin}\ and\ \citenamefont
  {Piran}(2001)}]{Sorkin:2000}%
  \BibitemOpen
  \bibfield  {author} {\bibinfo {author} {\bibfnamefont {E.}~\bibnamefont
  {Sorkin}}\ and\ \bibinfo {author} {\bibfnamefont {T.}~\bibnamefont {Piran}},\
  }\href {\doibase 10.1103/PhysRevD.63.084006} {\bibfield  {journal} {\bibinfo
  {journal} {Phys. Rev. D}\ }\textbf {\bibinfo {volume} {63}},\ \bibinfo
  {pages} {084006} (\bibinfo {year} {2001})},\ \Eprint
  {http://arxiv.org/abs/gr-qc/0009095} {arXiv:gr-qc/0009095} \BibitemShut
  {NoStop}%
\bibitem [{\citenamefont {Schwinger}(1951)}]{Schwinger:1951}%
  \BibitemOpen
  \bibfield  {author} {\bibinfo {author} {\bibfnamefont {J.~S.}\ \bibnamefont
  {Schwinger}},\ }\href {\doibase 10.1103/PhysRev.82.664} {\bibfield  {journal}
  {\bibinfo  {journal} {Phys. Rev.}\ }\textbf {\bibinfo {volume} {82}},\
  \bibinfo {pages} {664} (\bibinfo {year} {1951})}\BibitemShut {NoStop}%
\bibitem [{\citenamefont {Klein}\ and\ \citenamefont
  {Zahn}(2021)}]{Klein:2021}%
  \BibitemOpen
  \bibfield  {author} {\bibinfo {author} {\bibfnamefont {C.}~\bibnamefont
  {Klein}}\ and\ \bibinfo {author} {\bibfnamefont {J.}~\bibnamefont {Zahn}},\
  }\href {\doibase 10.1103/PhysRevD.104.025009} {\bibfield  {journal} {\bibinfo
   {journal} {Phys. Rev. D}\ }\textbf {\bibinfo {volume} {104}},\ \bibinfo
  {pages} {025009} (\bibinfo {year} {2021})},\ \Eprint
  {http://arxiv.org/abs/2104.06005} {arXiv:2104.06005 [gr-qc]} \BibitemShut
  {NoStop}%
\bibitem [{\citenamefont {Klein}\ \emph {et~al.}(2021)\citenamefont {Klein},
  \citenamefont {Zahn},\ and\ \citenamefont {Hollands}}]{Klein:2021a}%
  \BibitemOpen
  \bibfield  {author} {\bibinfo {author} {\bibfnamefont {C.}~\bibnamefont
  {Klein}}, \bibinfo {author} {\bibfnamefont {J.}~\bibnamefont {Zahn}}, \ and\
  \bibinfo {author} {\bibfnamefont {S.}~\bibnamefont {Hollands}},\ }\href
  {\doibase 10.1103/PhysRevLett.127.231301} {\bibfield  {journal} {\bibinfo
  {journal} {Phys. Rev. Lett.}\ }\textbf {\bibinfo {volume} {127}},\ \bibinfo
  {pages} {231301} (\bibinfo {year} {2021})},\ \Eprint
  {http://arxiv.org/abs/2103.03714} {arXiv:2103.03714 [gr-qc]} \BibitemShut
  {NoStop}%
\bibitem [{\citenamefont {Zilberman}\ \emph
  {et~al.}(2022{\natexlab{a}})\citenamefont {Zilberman}, \citenamefont
  {Casals}, \citenamefont {Ori},\ and\ \citenamefont
  {Ottewill}}]{Zilberman:2022a}%
  \BibitemOpen
  \bibfield  {author} {\bibinfo {author} {\bibfnamefont {N.}~\bibnamefont
  {Zilberman}}, \bibinfo {author} {\bibfnamefont {M.}~\bibnamefont {Casals}},
  \bibinfo {author} {\bibfnamefont {A.}~\bibnamefont {Ori}}, \ and\ \bibinfo
  {author} {\bibfnamefont {A.~C.}\ \bibnamefont {Ottewill}},\ }\href {\doibase
  10.1103/PhysRevLett.129.261102} {\bibfield  {journal} {\bibinfo  {journal}
  {Phys. Rev. Lett.}\ }\textbf {\bibinfo {volume} {129}},\ \bibinfo {pages}
  {261102} (\bibinfo {year} {2022}{\natexlab{a}})},\ \Eprint
  {http://arxiv.org/abs/2203.08502} {arXiv:2203.08502 [gr-qc]} \BibitemShut
  {NoStop}%
\bibitem [{\citenamefont {Hintz}\ and\ \citenamefont
  {Vasy}(2017)}]{Hintz:2015}%
  \BibitemOpen
  \bibfield  {author} {\bibinfo {author} {\bibfnamefont {P.}~\bibnamefont
  {Hintz}}\ and\ \bibinfo {author} {\bibfnamefont {A.}~\bibnamefont {Vasy}},\
  }\href {\doibase 10.1063/1.4996575} {\bibfield  {journal} {\bibinfo
  {journal} {J.\ Math.\ Phys.}\ }\textbf {\bibinfo {volume} {58}},\ \bibinfo
  {pages} {081509} (\bibinfo {year} {2017})},\ \Eprint
  {http://arxiv.org/abs/1512.08004} {arXiv:1512.08004 [math.AP]} \BibitemShut
  {NoStop}%
\bibitem [{\citenamefont {Dyatlov}(2011)}]{Dyatlov:2010}%
  \BibitemOpen
  \bibfield  {author} {\bibinfo {author} {\bibfnamefont {S.}~\bibnamefont
  {Dyatlov}},\ }\href {\doibase 10.1007/s00220-011-1286-x} {\bibfield
  {journal} {\bibinfo  {journal} {Commun. Math. Phys.}\ }\textbf {\bibinfo
  {volume} {306}},\ \bibinfo {pages} {119} (\bibinfo {year} {2011})},\ \Eprint
  {http://arxiv.org/abs/1003.6128} {arXiv:1003.6128 [math.AP]} \BibitemShut
  {NoStop}%
\bibitem [{\citenamefont {Hintz}(2021)}]{Hintz:2021}%
  \BibitemOpen
  \bibfield  {author} {\bibinfo {author} {\bibfnamefont {P.}~\bibnamefont
  {Hintz}},\ }\href@noop {} {\  (\bibinfo {year} {2021})},\ \Eprint
  {http://arxiv.org/abs/2112.14431} {arXiv:2112.14431 [gr-qc]} \BibitemShut
  {NoStop}%
  \bibitem [{\citenamefont {Hintz}\ and\ \citenamefont
  {Klein}(2023)}]{KleinHintz}%
  \BibitemOpen
  \bibfield  {author} {\bibinfo {author} {\bibfnamefont {P.}~\bibnamefont
  {Hintz}}\ and\ \bibinfo {author} {\bibfnamefont {C.}~\bibnamefont {Klein}},\
  }\href@noop {} {\bibfield  {journal} {\bibinfo  {journal} {Class. Quant. Grav.}\ }\textbf {\bibinfo
  {volume} {41}},\ \bibinfo {pages} {075006} (\bibinfo {year} {2024})},\ \Eprint  
  {http://arxiv.org/abs/2310.19655} {arXiv:2310.19655 [gr-qc]}  \BibitemShut {NoStop}
\bibitem [{\citenamefont {Borthwick}(2018)}]{Borthwick:2018}%
  \BibitemOpen
  \bibfield  {author} {\bibinfo {author} {\bibfnamefont {J.}~\bibnamefont
  {Borthwick}},\ }\href {\doibase 10.1088/1361-6382/aae3dc} {\bibfield
  {journal} {\bibinfo  {journal} {Class. Quant. Grav.}\ }\textbf {\bibinfo
  {volume} {35}},\ \bibinfo {pages} {215006} (\bibinfo {year} {2018})},\
  \bibinfo {note} {[Erratum: Class.Quant.Grav. 39, 219501 (2022)]},\ \Eprint
  {http://arxiv.org/abs/1805.00243} {arXiv:1805.00243 [gr-qc]} \BibitemShut
  {NoStop}%
\bibitem [{\citenamefont {Klein}(2023)}]{Klein:2022}%
  \BibitemOpen
  \bibfield  {author} {\bibinfo {author} {\bibfnamefont {C.~K.~M.}\
  \bibnamefont {Klein}},\ }\href {\doibase 10.1007/s00023-023-01273-6}
  {\bibfield  {journal} {\bibinfo  {journal} {Annales Henri Poincare}\ }\textbf
  {\bibinfo {volume} {24}},\ \bibinfo {pages} {2401} (\bibinfo {year}
  {2023})},\ \Eprint {http://arxiv.org/abs/2206.05073} {arXiv:2206.05073
  [gr-qc]} \BibitemShut {NoStop}%
\bibitem [{\citenamefont {Wald}(1984)}]{Wald:1984}%
  \BibitemOpen
  \bibfield  {author} {\bibinfo {author} {\bibfnamefont {R.~M.}\ \bibnamefont
  {Wald}},\ }\href {\doibase 10.7208/chicago/9780226870373.001.0001} {\emph
  {\bibinfo {title} {{General Relativity}}}}\ (\bibinfo  {publisher} {Chicago
  Univ. Pr.},\ \bibinfo {address} {Chicago, USA},\ \bibinfo {year}
  {1984})\BibitemShut {NoStop}%
\bibitem [{\citenamefont {Suzuki}\ \emph {et~al.}(1998)\citenamefont {Suzuki},
  \citenamefont {Takasugi},\ and\ \citenamefont {Umetsu}}]{Suzuki:1998}%
  \BibitemOpen
  \bibfield  {author} {\bibinfo {author} {\bibfnamefont {H.}~\bibnamefont
  {Suzuki}}, \bibinfo {author} {\bibfnamefont {E.}~\bibnamefont {Takasugi}}, \
  and\ \bibinfo {author} {\bibfnamefont {H.}~\bibnamefont {Umetsu}},\ }\href
  {\doibase 10.1143/PTP.100.491} {\bibfield  {journal} {\bibinfo  {journal}
  {Prog. Theor. Phys.}\ }\textbf {\bibinfo {volume} {100}},\ \bibinfo {pages}
  {491} (\bibinfo {year} {1998})},\ \Eprint
  {http://arxiv.org/abs/gr-qc/9805064} {arXiv:gr-qc/9805064} \BibitemShut
  {NoStop}%
\bibitem [{\citenamefont {Suzuki}\ \emph {et~al.}(1999)\citenamefont {Suzuki},
  \citenamefont {Takasugi},\ and\ \citenamefont {Umetsu}}]{Suzuki:1999}%
  \BibitemOpen
  \bibfield  {author} {\bibinfo {author} {\bibfnamefont {H.}~\bibnamefont
  {Suzuki}}, \bibinfo {author} {\bibfnamefont {E.}~\bibnamefont {Takasugi}}, \
  and\ \bibinfo {author} {\bibfnamefont {H.}~\bibnamefont {Umetsu}},\ }\href
  {\doibase 10.1143/PTP.102.253} {\bibfield  {journal} {\bibinfo  {journal}
  {Prog. Theor. Phys.}\ }\textbf {\bibinfo {volume} {102}},\ \bibinfo {pages}
  {253} (\bibinfo {year} {1999})},\ \Eprint
  {http://arxiv.org/abs/gr-qc/9905040} {arXiv:gr-qc/9905040} \BibitemShut
  {NoStop}%
\bibitem [{\citenamefont {Zilberman}\ \emph
  {et~al.}(2022{\natexlab{b}})\citenamefont {Zilberman}, \citenamefont
  {Casals}, \citenamefont {Ori},\ and\ \citenamefont
  {Ottewill}}]{Zilberman:2022}%
  \BibitemOpen
  \bibfield  {author} {\bibinfo {author} {\bibfnamefont {N.}~\bibnamefont
  {Zilberman}}, \bibinfo {author} {\bibfnamefont {M.}~\bibnamefont {Casals}},
  \bibinfo {author} {\bibfnamefont {A.}~\bibnamefont {Ori}}, \ and\ \bibinfo
  {author} {\bibfnamefont {A.~C.}\ \bibnamefont {Ottewill}},\ }\href {\doibase
  10.1103/PhysRevD.106.125011} {\bibfield  {journal} {\bibinfo  {journal}
  {Phys. Rev. D}\ }\textbf {\bibinfo {volume} {106}},\ \bibinfo {pages}
  {125011} (\bibinfo {year} {2022}{\natexlab{b}})},\ \Eprint
  {http://arxiv.org/abs/2203.07780} {arXiv:2203.07780 [gr-qc]} \BibitemShut
  {NoStop}%
\bibitem [{\citenamefont {Hollands}\ and\ \citenamefont
  {Wald}(2001)}]{Hollands:2001nf}%
  \BibitemOpen
  \bibfield  {author} {\bibinfo {author} {\bibfnamefont {S.}~\bibnamefont
  {Hollands}}\ and\ \bibinfo {author} {\bibfnamefont {R.~M.}\ \bibnamefont
  {Wald}},\ }\href {\doibase 10.1007/s002200100540} {\bibfield  {journal}
  {\bibinfo  {journal} {Commun. Math. Phys.}\ }\textbf {\bibinfo {volume}
  {223}},\ \bibinfo {pages} {289} (\bibinfo {year} {2001})},\ \Eprint
  {http://arxiv.org/abs/gr-qc/0103074} {arXiv:gr-qc/0103074} \BibitemShut
  {NoStop}%
\bibitem [{\citenamefont {Li}\ and\ \citenamefont {Van~de
  Moortel}(2023)}]{li2023kasner}%
  \BibitemOpen
  \bibfield  {author} {\bibinfo {author} {\bibfnamefont {W.}~\bibnamefont
  {Li}}\ and\ \bibinfo {author} {\bibfnamefont {M.}~\bibnamefont {Van~de
  Moortel}},\ }\href@noop {} {\bibfield  {journal} {\bibinfo  {journal} {arXiv
  preprint arXiv:2302.00046}\ } (\bibinfo {year} {2023})}\BibitemShut {NoStop}%
\bibitem [{\citenamefont {{Casals}}\ and\ \citenamefont
  {{Marinho}}(2022)}]{2022PhRvD.106d4060C}%
  \BibitemOpen
  \bibfield  {author} {\bibinfo {author} {\bibfnamefont {M.}~\bibnamefont
  {{Casals}}}\ and\ \bibinfo {author} {\bibfnamefont {C.~I.~S.}\ \bibnamefont
  {{Marinho}}},\ }\href {\doibase 10.1103/PhysRevD.106.044060} {\bibfield
  {journal} {\bibinfo  {journal} {Phys.Rev. D}\ }\textbf {\bibinfo {volume}
  {106}},\ \bibinfo {eid} {044060} (\bibinfo {year} {2022})},\ \Eprint
  {http://arxiv.org/abs/2006.06483} {arXiv:2006.06483 [gr-qc]} \BibitemShut
  {NoStop}%
\bibitem [{\citenamefont {{Casals}}\ and\ \citenamefont {{Teixeira da
  Costa}}(2022)}]{2022CMaPh.394..797C}%
  \BibitemOpen
  \bibfield  {author} {\bibinfo {author} {\bibfnamefont {M.}~\bibnamefont
  {{Casals}}}\ and\ \bibinfo {author} {\bibfnamefont {R.}~\bibnamefont
  {{Teixeira da Costa}}},\ }\href {\doibase 10.1007/s00220-022-04410-0}
  {\bibfield  {journal} {\bibinfo  {journal} {Communications in Mathematical
  Physics}\ }\textbf {\bibinfo {volume} {394}},\ \bibinfo {pages} {797}
  (\bibinfo {year} {2022})},\ \Eprint {http://arxiv.org/abs/2105.13329}
  {arXiv:2105.13329 [gr-qc]} \BibitemShut {NoStop}%
  \bibitem [{\citenamefont {Gregory}\ \emph {et~al.}(2021)\citenamefont
  {Gregory}, \citenamefont {Moss}, \citenamefont {Oshita},\ and\ \citenamefont
  {Patrick}}]{Gregory:2021}%
  \BibitemOpen
  \bibfield  {author} {\bibinfo {author} {\bibfnamefont {R.}~\bibnamefont
  {Gregory}}, \bibinfo {author} {\bibfnamefont {I.~G.}\ \bibnamefont {Moss}},
  \bibinfo {author} {\bibfnamefont {N.}~\bibnamefont {Oshita}}, \ and\ \bibinfo
  {author} {\bibfnamefont {S.}~\bibnamefont {Patrick}},\ }\href {\doibase
  10.1088/1361-6382/ac1a68} {\bibfield  {journal} {\bibinfo  {journal} {Class.
  Quant. Grav.}\ }\textbf {\bibinfo {volume} {38}},\ \bibinfo {pages} {185005}
  (\bibinfo {year} {2021})},\ \Eprint {http://arxiv.org/abs/2103.09862}
  {arXiv:2103.09862 [gr-qc]} \BibitemShut {NoStop}%
  \bibitem [{\citenamefont {Hatsuda}(2020)}]{Hatsuda:2020}%
  \BibitemOpen
  \bibfield  {author} {\bibinfo {author} {\bibfnamefont {Y.}~\bibnamefont
  {Hatsuda}},\ }\href {\doibase 10.1088/1361-6382/abc82e} {\bibfield  {journal}
  {\bibinfo  {journal} {Class. Quant. Grav.}\ }\textbf {\bibinfo {volume}
  {38}},\ \bibinfo {pages} {025015} (\bibinfo {year} {2020})},\ \Eprint
  {http://arxiv.org/abs/2006.08957} {arXiv:2006.08957 [gr-qc]} \BibitemShut
  {NoStop}%
\end{thebibliography}
%


\end{document}